\documentclass[showpacs,preprintnumbers,twocolumn,superscriptaddress,aps,nofootinbib]{revtex4-1}
\usepackage{amssymb}
\usepackage[centertags]{amsmath}
\usepackage{txfonts}
\usepackage{epsfig}
\usepackage{bm}
\usepackage{color}
\usepackage{graphicx,graphics}
\usepackage{multirow}
\usepackage{float}
\usepackage{ulem}
\usepackage{appendix}
\usepackage[pdfstartview=FitH]{hyperref}
\hypersetup{colorlinks=true, citecolor=blue,linkcolor=blue,filecolor=black,urlcolor=blue}  
\allowdisplaybreaks[2]

\begin{document}

\title{Productions of $^3_{\Lambda}$H, $^4_{\Lambda}$H and $^4_{\Lambda}$He in different coalescence channels in Au-Au collisions at $\sqrt{s_{NN}}=3$ GeV}

\author{Rui-Qin Wang}
\affiliation{School of Physics and Physical Engineering, Qufu Normal University, Shandong 273165, China}

\author{Jun Song}
\affiliation{School of Physical Science and Intelligent Engineering, Jining University, Shandong 273155, China}

\author{Mao-Yan Wu}
\affiliation{School of Physics and Physical Engineering, Qufu Normal University, Shandong 273165, China}

\author{Huai-Tong Xue}
\affiliation{College of Physics and Electronic Engineering, Nanyang Normal University, Nanyang 473061, China}

\author{Feng-Lan Shao}
\email {shaofl@mail.sdu.edu.cn}
\affiliation{School of Physics and Physical Engineering, Qufu Normal University, Shandong 273165, China}

\begin{abstract}

We study the productions of $\Lambda$-hypernuclei $^3_{\Lambda}$H, $^4_{\Lambda}$H and $^4_{\Lambda}$He in the coalescence mechanism in Au-Au collisions at $\sqrt{s_{NN}}=3$ GeV. 
Considering the abundance and great importance of baryons and light (hyper-)nuclei on the collision dynamics, we include not only nucleon$+\Lambda$ coalescence but also nucleus+nucleon($\Lambda$) coalescence.
We present contributions from different coalescence channels for $^3_{\Lambda}$H, $^4_{\Lambda}$H and $^4_{\Lambda}$He in their productions.
We predict the production asymmetry between $^4_{\Lambda}$H and $^4_{\Lambda}$He, characterized by yield ratios $^4_{\Lambda}\text{He}/^4_{\Lambda}\text{H}$ and $(^4_{\Lambda}\text{H}-^4_{\Lambda}\text{He})/(^4_{\Lambda}\text{H}+^4_{\Lambda}\text{He})$, which can shed light on the existence constraints of the possible neutron-$\Lambda$ bound states $^2_{\Lambda}n~(n\Lambda)$ and $^3_{\Lambda}n~(nn\Lambda)$.

\end{abstract}

\maketitle

\section{Introduction}

Hypernuclei, nuclear systems with one or more constituent nucleons ($p$ and $n$) replaced by hyperons, are a special group of observables in relativistic heavy ion collisions.
They serve as natural tools for the study of elementary hyperon-nucleon and hyperon-hyperon interactions~\cite{Chen:2018tnh,Saito:2021gao}.
Due to their small binding energies, hypernuclei are considered to be formed at the late stage of the system evolution and so they are good probes of the fireball freeze-out properties~\cite{Andronic:2017pug,Kozhevnikova:2024itb}.
Investigations into hypernuclei deal with several aspects, mainly including production, decay, internal structure, and spectroscopy~\cite{Gal:2016boi,ExHIC:2017smd,Hashimoto:2006aw,Feliciello:2015dua}. 

Productions of hypernuclei in relativistic heavy ion collisions have attracted much attention in experiment~\cite{Davis:2005mb,Esser:2013aya,Chen:2024aom} and theory~\cite{Nemura:1999qp,Steinheimer:2012tb,Botvina:2016wko,Bertulani:2022vad,Reichert:2022mek}.
Several experiments, such as the STAR at the BNL Relativistic Heavy Ion Collider (RHIC)~\cite{STAR:2019wjm,STAR:2021orx,STAR:2022fnj,STAR:2023fbc} and the ALICE at the CERN Large Hadron Collider (LHC)~\cite{ALICE:2022sco,ALICE:2015oer,ALICE:2024koa,ALICE:2024ilx}, have collected a relative wealth of data on hypernucleus formations.
In theory, two popular production mechanisms, the thermal production mechanism~\cite{Andronic:2010qu,Cleymans:2011pe,Vovchenko:2018fiy} and the coalescence mechanism~\cite{Liu:2024ygk,Sun:2018mqq,Sun:2016rev}, have been employed to successfully describe some production properties of hypernuclei. 

In recent years, the STAR collaboration has extended the measurements of hypernuclei up to the mass number $A=4$ in Au-Au collisions at $\sqrt{s_{NN}}=3$ GeV~\cite{STAR:2021orx,STAR:2022fnj}.
At such relatively low collision energy, baryonic interactions dominate the collision dynamics~\cite{STAR:2021ozh}, and many different characteristics compared to those at high RHIC and LHC energies appear, such as the disappearance of the partonic collectivity~\cite{STAR:2021yiu}.
In spite of this, both the first-order azimuthal anisotropies (i.e., the directed flows) and the averaged transverse momenta of different species of light nuclei and hypernuclei experimentally measured at midrapidity area still follow the mass number scaling~\cite{STAR:2021ozh,STAR:2023uxk,STAR:2022fnj}.
This indicates the validity of the hadronic collectivity and the dominance of the coalescence production mechanism for these composite clusters.
Different event generators implanted the coalescence mechanism have been used to explain these recently-measured data on light nuclei and hypernuclei from different viewpoints~\cite{Kozhevnikova:2024itb,Liu:2024ilw,Xu:2023xul}.

In order to intuitively see whether there are new characteristics and which of them come from the coalescence mechanism itself, we, in our previous work~\cite{Wang:2022hja}, developed an analytical description for the productions of different species of light nuclei (from $A=2$ to $A=4$) in Au-Au collisions at $\sqrt{s_{NN}}=3$ GeV.
As light nuclei are abundantly produced in heavy-ion collisions at such relatively low energy~\cite{STAR:2023uxk}, they are expected to have significant effects on the collision dynamics, and it requires treating light nuclei on the same footing as nucleons~\cite{Wang:2023gta}.
Based on this point, we introduced nucleus+nucleon coalescence besides nucleon+nucleon coalescence and found it played a requisite role~\cite{Wang:2022hja}.
For productions of hypernuclei with added strangeness degree of freedom, besides nucleon+hyperon coalescence, are coalescence processes involving nucleus participation still necessary?
If yes, what new characteristics will they bring?

In this work, we extend the coalescence model in Ref.~\cite{Wang:2022hja} to the strange sector, including not only nucleon$+\Lambda$ coalescence but also nucleus+nucleon($\Lambda$) coalescence, and use it to study the productions of $\Lambda$-hypernuclei $^3_{\Lambda}$H, $^4_{\Lambda}$H and $^4_{\Lambda}$He in Au-Au collisions at $\sqrt{s_{NN}}=3$ GeV.
We present contributions from different coalescence channels for $^3_{\Lambda}$H, $^4_{\Lambda}$H and $^4_{\Lambda}$He.
Compared with the available data of $^4_{\Lambda}$H measured by the STAR collaboration, we find besides the deuteron ($d$) and the triton ($t$), another two possible neutron-$\Lambda$ bound states $^2_{\Lambda}n$ and $^3_{\Lambda}n$ could also take part in the $^4_{\Lambda}$H formation.
If they indeed exist, $^2_{\Lambda}n$ can enter into $^4_{\Lambda}$H via two different coalescence channels $p+n+^2_{\Lambda}n$ and $d+^2_{\Lambda}n$, but only one into $^4_{\Lambda}$He via $p+p+^2_{\Lambda}n$.
For $^3_{\Lambda}n$, it could only enter into $^4_{\Lambda}$H.
So the existences of $^2_{\Lambda}n$ and $^3_{\Lambda}n$ can enhance $^4_{\Lambda}$H production much stronger than $^4_{\Lambda}$He, and then enlarge the production asymmetry between $^4_{\Lambda}$H and $^4_{\Lambda}$He.
We propose the yield ratios $^4_{\Lambda}\text{He}/^4_{\Lambda}\text{H}$ and $(^4_{\Lambda}\text{H}-^4_{\Lambda}\text{He})/(^4_{\Lambda}\text{H}+^4_{\Lambda}\text{He})$ to characterize such asymmetry and give the corresponding predictions.
Comparisons with future measurements can help shed light on the existence constraints of $^2_{\Lambda}n$ and $^3_{\Lambda}n$.

The rest of the paper is organized as follows. 
In Sec.~\ref{model}, we introduce the coalescence model. 
We present analytical formulas of momentum distributions of light (hyper-)nuclei formed via different coalescence channels.
In Sec.~\ref{results}, we apply the coalescence model to Au-Au collisions at $\sqrt{s_{NN}}=3$ GeV to study the productions of $^3_{\Lambda}$H, $^4_{\Lambda}$H and $^4_{\Lambda}$He.
We give results of $^3_{\Lambda}$H, $^4_{\Lambda}$H and $^4_{\Lambda}$He in different cases, with and without existences of $^2_{\Lambda}n$ and $^3_{\Lambda}n$.
In Sec.~\ref{summary}, we summarize our work.

\section{The coalescence model}   \label{model}

We extend the coalescence model in our previous work~\cite{Wang:2022hja} to manage the productions of hypernuclei besides light nuclei.
The model's starting point is a hadronic system formed at the late stage of the evolution of high-energy heavy ion collision.
The hadronic system consists of different species of primordial mesons, nucleons, hyperons, and their antiparticles.
Our model employs a two-step method to consider all possible coalescence channels resulting in light (hyper-)nucleus production.
In the first step, all primordial nucleons and hyperons are allowed to form $d$, $t$, $^3$He, $^3_{\Lambda}$H, $^4_{\Lambda}$H and so on via the nucleon+nucleon/hyperon coalescence.
Then in the second step, the previously formed nuclei such as $d$ and $^3$He capture the remanent primordial nucleons and/or hyperons, i.e., those excluding consumed ones in the first step, to recombine into heavier nuclei with larger mass numbers.

Such a two-step method includes not only the baryon interactions resulting in the light (hyper-)nucleus production, but also the interactions between the previously formed nuclei and the remanent baryons.
What's more, there is no double-counting in such a method.
Take the triton production as an example to illustrate this point.
We stamp the protons and neutrons forming deuterons $p_a,n_a$, those forming tritons $p_b,n_b$ and so on in the first step.
For the remanent nucleons after the first step, we stamp them with $p_{rem},n_{rem}$.
Then in the second step, we judge the previously formed deuterons.
If they and the remanent neutrons meet coalescence requirements, they capture the remanent neutrons to merge into tritons. 
Therefore, if the triton is formed via the channel $p+n+n\rightarrow t$, the consumed nucleons are from those $p_b,n_b$'s,
and if the triton is formed via the channel $d+n\rightarrow t$, the consumed nucleons are those $p_a,n_a$'s and $p_{rem},n_{rem}$'s.
This explicitly shows there is no double-counting in different coalescence channels in the two-step method.

In the following, we present the deduction of the production formalism of light (hyper-)nuclei via different coalescence processes. 
First, we give analytical results of two bodies coalescing into light (hyper-)nuclei, which can be applied to processes such as $p+n \rightarrow d$, $d+\Lambda \rightarrow ^3_{\Lambda}$H, and $t+\Lambda \rightarrow ^4_{\Lambda}$H.
Then we show analytical results of three bodies coalescing into light (hyper-)nuclei, which can be used to describe these processes, e.g., $p+n+\Lambda \rightarrow ^3_{\Lambda}$H, and $n+d+\Lambda \rightarrow ^4_{\Lambda}$H.
We also show the analytical result of four bodies coalescing into light (hyper-)nuclei, i.e., $p+n+n+\Lambda \rightarrow ^4_{\Lambda}$H and $p+p+n+\Lambda \rightarrow ^4_{\Lambda}$He. 
Finally, we give the general formalism of $N$-body coalescence.

\subsection{Formalism of two-body coalescence}     \label{2hcoHj}

In the framework of two hadronic bodies $h_1$ and $h_2$ coalescing into light (hyper-)nuclei $H_j$, the three-dimensional momentum distribution $f_{H_j}(\bm{p})$ of $H_j$ is given by
{\setlength\arraycolsep{0pt}
\begin{eqnarray}
 f_{H_j}(\bm{p}) =&&  \int d\bm{x}_1d\bm{x}_2 d\bm{p}_1 d\bm{p}_2  f_{h_1h_2}(\bm{x}_1,\bm{x}_2;\bm{p}_1,\bm{p}_2)  \nonumber  \\
 &&~~~~ \times \mathcal {R}_{H_j}(\bm{x}_1,\bm{x}_2;\bm{p}_1,\bm{p}_2,\bm{p}),      \label{eq:fHj2hgeneral} 
\end{eqnarray} }%
where $f_{h_1h_2}(\bm{x}_1,\bm{x}_2;\bm{p}_1,\bm{p}_2)$ is two-hadron joint coordinate-momentum distribution; 
$\mathcal {R}_{H_j}(\bm{x}_1,\bm{x}_2;\bm{p}_1,\bm{p}_2,\bm{p})$ is the kernel function.
Here and from now on, we use bold symbols to denote three-dimensional coordinates or momenta.

In terms of the normalized joint coordinate-momentum distribution denoted by the superscript `$(n)$', we have
{\setlength\arraycolsep{0pt}
\begin{eqnarray}
 f_{H_j}(\bm{p})=&& N_{h_1h_2} \int d\bm{x}_1d\bm{x}_2 d\bm{p}_1 d\bm{p}_2  f^{(n)}_{h_1h_2}(\bm{x}_1,\bm{x}_2;\bm{p}_1,\bm{p}_2)  \nonumber \\
  && ~~~~~~~~~~~~~ \times \mathcal {R}_{H_j}(\bm{x}_1,\bm{x}_2;\bm{p}_1,\bm{p}_2,\bm{p}).      \label{eq:fHj2hgeneral1}               
\end{eqnarray} }%
$N_{h_1h_2}$ is the number of all possible $h_1h_2$-pairs, and it is equal to $N_{h_1}N_{h_2}$.
$N_{h_i}~ (i=1,2)$ is the number of the hadrons $h_i$ in the considered hadronic system.

The kernel function $\mathcal {R}_{H_j}(\bm{x}_1,\bm{x}_2;\bm{p}_1,\bm{p}_2,\bm{p})$ denotes the probability density for $h_1$, $h_2$ with
momenta $\bm{p}_1$ and $\bm{p}_2$ at $\bm{x}_1$ and $\bm{x}_2$ to combine into an $H_j$ of momentum $\bm{p}$.
It carries the kinetic and dynamical information of $h_1$ and $h_2$ combining process,
and its precise expression should be constrained by such as the momentum conservation and constraints due to intrinsic quantum numbers, e.g. spin, and so on.
To explicitly take these constraints into account, we rewrite the kernel function as in Refs.~\cite{Wang:2020zaw,Zhao:2022xkz} in the following form
{\setlength\arraycolsep{0pt}
\begin{eqnarray}
  \mathcal {R}_{H_j}(\bm{x}_1,\bm{x}_2;\bm{p}_1,\bm{p}_2,\bm{p}) =&& g_{H_j} \mathcal {R}_{H_j}^{(x,p)}(\bm{x}_1,\bm{x}_2;\bm{p}_1,\bm{p}_2) \nonumber \\  
 && \times \delta(\displaystyle{\sum^2_{i=1}} \bm{p}_i-\bm{p}).     \label{eq:RHj2h}  
\end{eqnarray} }%
The spin degeneracy factor $g_{H_j} = (2J_{H_j}+1) /[\prod \limits_{i=1}^2(2J_{h_i}+1)]$.
$J_{H_j}$ is the spin of the produced $H_j$ and $J_{h_i}$ is that of the primordial hadron $h_i$. 
The Dirac $\delta$ function guarantees the momentum conservation in the coalescence.
The remaining $\mathcal {R}_{H_j}^{(x,p)}(\bm{x}_1,\bm{x}_2;\bm{p}_1,\bm{p}_2)$ can be solved from the Wigner transformation once the wave function of $H_j$ is given with the instantaneous coalescence approximation.
It is as follows 
{\setlength\arraycolsep{0pt}
\begin{eqnarray}
&&  \mathcal {R}^{(x,p)}_{H_j}(\bm{x}_1,\bm{x}_2;\bm{p}_1,\bm{p}_2) = 8e^{-\frac{(\bm{x}'_1-\bm{x}'_2)^2}{2\sigma_{21}^2}}
     e^{-\frac{2\sigma_{21}^2(m_2\bm{p}'_{1}-m_1\bm{p}'_{2})^2}{(m_1+m_2)^2}},      \label{eq:RHj2hxp} 
\end{eqnarray} }%
as we adopt the wave function of a spherical harmonic oscillator~\cite{Chen:2003ava,Zhu:2015voa}.
The superscript `$'$' in the coordinate or momentum variable denotes the hadronic coordinate or momentum in the rest frame of the $h_1h_2$-pair.
$m_1$ and $m_2$ are the mass of hadron $h_1$ and that of hadron $h_2$.
The width parameter $\sigma_{21}=\sqrt{\frac{2(m_1+m_2)^2}{3(m_1^2+m_2^2)}} RMS_{H_j}$,
where $RMS_{H_j}$ is the root-mean-square radius of $H_j$ and its values for different light nuclei can be found in Ref.~\cite{Angeli:2013epw}.

The normalized two-hadron joint distribution $f^{(n)}_{h_1h_2}(\bm{x}_1,\bm{x}_2;\bm{p}_1,\bm{p}_2)$ is generally coordinate and momentum coupled, 
especially in central heavy-ion collisions with relatively high collision energies where the collective expansion is long-lasting. 
The coupling intensities and specific forms are probably different at different phase spaces in different collision energies and different collision centralities. 
Such a coupling effect on the production properties of light nuclei in Pb-Pb collisions at LHC has been studied in our recent work~\cite{Wang:2023rpd}.
At relatively low RHIC energies $\sqrt{s_{NN}}=7.7-54.4$ GeV, the coalescence model ignoring this coordinate and momentum coupling has proved to be successful in explaining the experimental data of $d$, $\bar d$ and $t$~\cite{Zhao:2022xkz}.
In this article, we try our best to derive production formulas analytically and more intuitively in Au-Au collisions at lower RHIC energy $\sqrt{s_{NN}}=3$ GeV where the partonic collectivity disappears~\cite{STAR:2021yiu}, so we consider a simple case that the joint distribution is coordinate and momentum factorized, i.e.,
{\setlength\arraycolsep{0pt}
\begin{eqnarray}
&& f^{(n)}_{h_1h_2}(\bm{x}_1,\bm{x}_2;\bm{p}_1,\bm{p}_2) = f^{(n)}_{h_1h_2}(\bm{x}_1,\bm{x}_2)   f^{(n)}_{h_1h_2}(\bm{p}_1,\bm{p}_2).  \label{eq:fh1h2fac}   
\end{eqnarray} }%

Substituting Eqs.~(\ref{eq:RHj2h}-\ref{eq:fh1h2fac}) into Eq.~(\ref{eq:fHj2hgeneral1}), we have
{\setlength\arraycolsep{0.2pt}
\begin{eqnarray}
&& f_{H_j}(\bm{p})= N_{h_1h_2} g_{H_j} \int d\bm{x}_1d\bm{x}_2 f^{(n)}_{h_1h_2}(\bm{x}_1,\bm{x}_2)  8e^{-\frac{(\bm{x}'_1-\bm{x}'_2)^2}{2\sigma_{21}^2}}  \nonumber   \\
&&~~~~~~ \times
 \int d\bm{p}_1d\bm{p}_2 f^{(n)}_{h_1h_2}(\bm{p}_1,\bm{p}_2)  e^{-\frac{2\sigma_{21}^2(m_2\bm{p}'_{1}-m_1\bm{p}'_{2})^2}{(m_1+m_2)^2}}
   \delta(\displaystyle{\sum^2_{i=1}} \bm{p}_i-\bm{p})   \nonumber   \\
&&~~~~~~~~~ = N_{h_1h_2} g_{H_j} \mathcal {A}_{H_j}  \mathcal {M}_{H_j}(\bm{p}),    \label{eq:fHj2h}  
\end{eqnarray} }%
where we use $\mathcal {A}_{H_j}$ to denote the coordinate integral part as
{\setlength\arraycolsep{0pt}
\begin{eqnarray}
\mathcal {A}_{H_j} =  8\int d\bm{x}_1d\bm{x}_2 f^{(n)}_{h_1h_2}(\bm{x}_1,\bm{x}_2) e^{-\frac{(\bm{x}'_1-\bm{x}'_2)^2}{2\sigma_{21}^2}},      \label{eq:AHj2h}      
\end{eqnarray} }%
and use $\mathcal {M}_{H_j}(\bm{p})$ to denote the momentum integral part as
{\setlength\arraycolsep{0pt}
\begin{eqnarray}
 \mathcal {M}_{H_j}(\bm{p}) = 
    \int d\bm{p}_1d\bm{p}_2 f^{(n)}_{h_1h_2}(\bm{p}_1,\bm{p}_2)  e^{-\frac{2\sigma_{21}^2(m_2\bm{p}'_{1}-m_1\bm{p}'_{2})^2}{(m_1+m_2)^2}}
   \delta(\displaystyle{\sum^2_{i=1}} \bm{p}_i-\bm{p}).  \nonumber \\   \label{eq:MHj2h}
\end{eqnarray} }%
Here $\mathcal {A}_{H_j}$ stands for the probability of an $h_1h_2$-pair satisfying the coordinate requirement to recombine into $H_j$,
and $\mathcal {M}_{H_j}(\bm{p})$ stands for the probability density of an $h_1h_2$-pair satisfying the momentum requirement to recombine into $H_j$ with momentum $\bm{p}$.

Changing integral variables to be $\bm{X}= \frac{\sqrt{2}(m_1\bm{x}_1+m_2\bm{x}_2)}{m_1+m_2}$ and $\bm{r}= \frac{ \bm{x}_1-\bm{x}_2}{\sqrt{2}}$ in Eq.~(\ref{eq:AHj2h}), we have
{\setlength\arraycolsep{0pt}
\begin{eqnarray}
 \mathcal {A}_{H_j} =  8\int d\bm{X} d\bm{r} f^{(n)}_{h_1h_2}(\bm{X},\bm{r}) e^{-\frac{\bm{r}'^2}{\sigma_{21}^2}},   \label{eq:AHj2hXr}   
\end{eqnarray} }%
and the normalizing condition
{\setlength\arraycolsep{0pt}
\begin{eqnarray}
 \int f^{(n)}_{h_1h_2}(\bm{X},\bm{r})  d\bm{X}d\bm{r}=1.
\end{eqnarray} }%
Considering the strong interaction and the coalescence are local, we further assume the center-of-mass coordinate is factorized, i.e., $f^{(n)}_{h_1h_2}(\bm{X},\bm{r}) = f^{(n)}_{h_1h_2}(\bm{X}) f^{(n)}_{h_1h_2}(\bm{r})$.
Adopting $f^{(n)}_{h_1h_2}(\bm{r}) = \frac{1}{(\pi C_1 R_f^2)^{3/2}} e^{-\frac{\bm{r}^2}{C_1 R_f^2}}$ as in Refs.~\cite{Mrowczynski:2016xqm,Wang:2020zaw}, we have
{\setlength\arraycolsep{0pt}
\begin{eqnarray}
&& \mathcal {A}_{H_j} =  \frac{8}{(\pi C_1 R_f^2)^{3/2}} \int d\bm{r} e^{-\frac{\bm{r}^2}{C_1R_f^2}} e^{-\frac{\bm{r}'^2}{\sigma_{21}^2}}.   \label{eq:AHj2h-r}  
\end{eqnarray} }%
$C_1$ is introduced to make $\bm{r}^2/C_1$ the square of one-half of the relative position, and it is 2~\cite{Mrowczynski:2016xqm,Kisiel:2014upa,ALICE:2015tra}.
In this way,  the effective radius of the hadronic system at the $H_j$ freeze-out $R_f$ is just the Hanbury-Brown-Twiss (HBT) interferometry radius, which can also be extracted from the two-particle femtoscopic correlations~\cite{Kisiel:2014upa,ALICE:2015tra}.

Considering instantaneous coalescence in the rest frame of $h_1h_2$-pair, i.e., $\Delta t'=0$, we get
\begin{eqnarray}
\bm{r} = \bm{r}' +(\gamma-1)\frac{\bm{r}'\cdot \bm{\beta}}{\beta^2}\bm{\beta},    \label{eq:LorentzTr}  
\end{eqnarray}
where $\bm{\beta}$ is the three-dimensional velocity of the center-of-mass frame of $h_1h_2$-pair in the laboratory frame and the Lorentz contraction factor $\gamma=1/\sqrt{1-\bm{\beta}^2}$.
The instantaneous coalescence is a basic assumption in coalescence-like models where the overlap of the nucleus Wigner phase-space density with the constituent phase-space distributions is adopted~\cite{Chen:2003ava}.
Considering the coalescence criterion judging in the rest frame is more general than in the laboratory frame, we choose the instantaneous coalescence in the rest frame of $h_1h_2$-pair, as in Refs.~\cite{Gutbrod:1976zzr,Chen:2003ava}.
Substituting Eq.~(\ref{eq:LorentzTr}) into Eq.~(\ref{eq:AHj2h-r}) and integrating from the relative coordinate variable, we can obtain 
{\setlength\arraycolsep{0pt}
\begin{eqnarray}
 \mathcal {A}_{H_j} = \frac{8\sigma_{21}^3}{(C_1 R_f^2+\sigma_{21}^2) \sqrt{C_1 (R_f/\gamma)^2+\sigma_{21}^2}}.   \label{eq:AHj2h-final}  
\end{eqnarray} }%

Noticing that $1/\sigma_{21}$ in Eq.~(\ref{eq:MHj2h}) has a small value of about or less than 0.1 GeV/c, we can mathematically approximate the gaussian form of the momentum-dependent kernel function to be a $\delta$ function as follows
\begin{equation}
e^{-\frac{(\bm{p}'_{1}-\frac{m_1}{m_2}\bm{p}'_{2})^2} {(1+\frac{m_1}{m_2})^2 /(2\sigma_{21}^2)}} \approx
\left[ \frac{ \sqrt{\pi}}{\sqrt{2}\sigma_{21}} (1+\frac{m_1}{m_2}) \right]^3 \delta(\bm{p}'_{1}-\frac{m_1}{m_2}\bm{p}'_{2}).
\end{equation}
The robustness of this $\delta$ function approximation has been checked at the outset of the analytical coalescence model in the work~\cite{Wang:2020zaw}.
After integrating $\bm{p}_1$ and $\bm{p}_2$ from Eq.~(\ref{eq:MHj2h}) we can obtain
{\setlength\arraycolsep{0.2pt}
\begin{eqnarray}
 \mathcal {M}_{H_{j}}(\bm{p}) =  (\frac{\sqrt{\pi}}{\sqrt{2}\sigma_{21}})^3 \gamma  f^{(n)}_{h_1h_2}(\frac{m_1\bm{p}}{m_1+m_2},\frac{m_2\bm{p}}{m_1+m_2}),  \label{eq:MHj2h-pfin}
\end{eqnarray} }%
where $\gamma$ comes from the momentum Lorentz transformation.

Substituting Eqs.~(\ref{eq:AHj2h-final}) and (\ref{eq:MHj2h-pfin}) into Eq.~(\ref{eq:fHj2h}) and ignoring correlations between $h_1$ and $h_2$ hadrons, we have
{\setlength\arraycolsep{0.2pt}
\begin{eqnarray}
 f_{H_j}(\bm{p})  &=& \frac{ 8\pi^{3/2} g_{H_j} \gamma}{ 2^{3/2} (C_1 R_f^2+\sigma_{21}^2) \sqrt{C_1 (R_f/\gamma)^2+\sigma_{21}^2}} 
                       f_{h_1}(\frac{m_1\bm{p}}{m_1+m_2})     \nonumber   \\
  &&  \times f_{h_2}(\frac{m_2\bm{p}}{m_1+m_2}).    \label{eq:fHj2h-approx}
\end{eqnarray} }%
Denoting the Lorentz invariant momentum distribution $\dfrac{d^{2}N}{2\pi p_{T}dp_{T}dy}$ with $f^{(\text{inv})}$, we finally have
{\setlength\arraycolsep{0.2pt}
\begin{eqnarray}
f_{H_j}^{(\text{inv})}(p_{T},y) &=&\frac{8 \pi^{3/2} g_{H_j} }{ 2^{3/2}(C_1 R_f^2+\sigma_{21}^2) \sqrt{C_1 (R_f/\gamma)^2+\sigma_{21}^2}}  \frac{m_{H_j}}{m_1m_2}   \nonumber   \\
  &&  \times        f_{h_1}^{(\text{inv})}(\frac{m_1p_{T}}{m_1+m_2},y)  f_{h_2}^{(\text{inv})}(\frac{m_2p_{T}}{m_1+m_2},y).     \label{eq:pt-Hj2h}
\end{eqnarray} }%
$p_T$ and $y$ are the transverse momentum and the longitudinal rapidity. $m_{H_j}$ is the mass of the formed $H_j$.

\subsection{Formalism of three-body coalescence}      \label{3hcoHj}

For light (hyper-)nuclei $H_j$ formed via the coalescence of three hadronic bodies $h_1$, $h_2$ and $h_3$, the three-dimensional momentum distribution $f_{H_j}(\bm{p})$ is
{\setlength\arraycolsep{0pt}
\begin{eqnarray}
 f_{H_j}(\bm{p})=&& N_{h_1h_2h_3} \int d\bm{x}_1d\bm{x}_2d\bm{x}_3 d\bm{p}_1 d\bm{p}_2d\bm{p}_3  f^{(n)}_{h_1h_2h_3}(\bm{x}_1,\bm{x}_2,\bm{x}_3;   \nonumber \\
  &&   \bm{p}_1,\bm{p}_2,\bm{p}_3)  \mathcal {R}_{H_j}(\bm{x}_1,\bm{x}_2,\bm{x}_3;\bm{p}_1,\bm{p}_2,\bm{p}_3,\bm{p}).     \label{eq:fHj3hgeneral1}               
\end{eqnarray} }%
$N_{h_1h_2h_3}$ is the number of all possible $h_1h_2h_3$-clusters.
It is equal to $N_{h_1}N_{h_2}N_{h_3}$ and $N_{h_1}(N_{h_1}-1)N_{h_3}$ for $h_1 \neq h_2 \neq h_3$ and $h_1 = h_2 \neq h_3$, respectively.
$f^{(n)}_{h_1h_2h_3}$ is the normalized three-hadron joint coordinate-momentum distribution.

We rewrite the kernel function as
{\setlength\arraycolsep{0pt}
\begin{eqnarray}
  \mathcal {R}_{H_j}(\bm{x}_1,\bm{x}_2,\bm{x}_3;\bm{p}_1,\bm{p}_2,\bm{p}_3,\bm{p}) =&& g_{H_j}
 \mathcal {R}_{H_j}^{(x,p)}(\bm{x}_1,\bm{x}_2,\bm{x}_3;\bm{p}_1,\bm{p}_2,\bm{p}_3) \nonumber \\  
 && \times \delta(\displaystyle{\sum^3_{i=1}} \bm{p}_i-\bm{p}).     \label{eq:RHj3h}  
\end{eqnarray} }%
The spin degeneracy factor $g_{H_j} = (2J_{H_j}+1) /[\prod \limits_{i=1}^3(2J_{h_i}+1)]$.
$\mathcal {R}_{H_j}^{(x,p)}(\bm{x}_1,\bm{x}_2,\bm{x}_3;\bm{p}_1,\bm{p}_2,\bm{p}_3)$ solving from the Wigner transformation~\cite{Chen:2003ava,Zhu:2015voa} is
{\setlength\arraycolsep{0pt}
\begin{eqnarray}
 && \mathcal {R}^{(x,p)}_{H_j}(\bm{x}_1,\bm{x}_2,\bm{x}_3;\bm{p}_1,\bm{p}_2,\bm{p}_3) = 8^2 e^{-\frac{(\bm{x}'_1-\bm{x}'_2)^2}{2\sigma_{31}^2}}
      e^{-\frac{2(\frac{m_1\bm{x}'_1}{m_1+m_2}+\frac{m_2\bm{x}'_2}{m_1+m_2}-\bm{x}'_3)^2}{3\sigma_{32}^2}}   \nonumber   \\
  &&~~~~~~~~~~~~ \times  e^{-\frac{2\sigma_{31}^2(m_2\bm{p}'_{1}-m_1\bm{p}'_{2})^2}{(m_1+m_2)^2}}
     e^{-\frac{3\sigma_{32}^2[m_3\bm{p}'_{1}+m_3\bm{p}'_{2}-(m_1+m_2)\bm{p}'_{3}]^2} {2(m_1+m_2+m_3)^2}}.      \label{eq:RHj3hxp} 
\end{eqnarray} }%
The superscript `$'$' denotes the coordinate or momentum in the rest frame of the $h_1h_2h_3$-cluster.
The width parameter $\sigma_{31}=\sqrt{\frac{m_3(m_1+m_2)(m_1+m_2+m_3)} {m_1m_2(m_1+m_2)+m_2m_3(m_2+m_3)+m_3m_1(m_3+m_1)}} RMS_{H_j}$,
and $\sigma_{32}=\sqrt{\frac{4m_1m_2(m_1+m_2+m_3)^2} {3(m_1+m_2)[m_1m_2(m_1+m_2)+m_2m_3(m_2+m_3)+m_3m_1(m_3+m_1)]}} RMS_{H_j}$.

With the coordinate and momentum factorization assumption of the joint distribution,
we have
{\setlength\arraycolsep{0.2pt}
\begin{eqnarray}
 f_{H_j}(\bm{p}) = N_{h_1h_2h_3} g_{H_j} \mathcal {A}_{H_j}  \mathcal {M}_{H_j}(\bm{p}).   \label{eq:fHj3h}  
\end{eqnarray} }%
Here we also use $\mathcal {A}_{H_j}$ to denote the coordinate integral part as
{\setlength\arraycolsep{0pt}
\begin{eqnarray}
\mathcal {A}_{H_j} = &&  8^2\int d\bm{x}_1d\bm{x}_2d\bm{x}_3 f^{(n)}_{h_1h_2h_3}(\bm{x}_1,\bm{x}_2,\bm{x}_3) e^{-\frac{(\bm{x}'_1-\bm{x}'_2)^2}{2\sigma_{31}^2}}   \nonumber  \\
&&~~~~~~~~ \times  e^{-\frac{2(\frac{m_1\bm{x}'_1}{m_1+m_2}+\frac{m_2\bm{x}'_2}{m_1+m_2}-\bm{x}'_3)^2}{3\sigma_{32}^2}} ,      \label{eq:AHj3h}      
\end{eqnarray} }%
and use $\mathcal {M}_{H_j}(\bm{p})$ to denote the momentum integral part as
{\setlength\arraycolsep{0pt}
\begin{eqnarray}
 \mathcal {M}_{H_j}(\bm{p}) = &&
    \int d\bm{p}_1d\bm{p}_2d\bm{p}_3 f^{(n)}_{h_1h_2h_3}(\bm{p}_1,\bm{p}_2,\bm{p}_3)   \delta(\displaystyle{\sum^3_{i=1}} \bm{p}_i-\bm{p})   \nonumber   \\
&&~~~~ \times    e^{-\frac{2\sigma_{31}^2(m_2\bm{p}'_{1}-m_1\bm{p}'_{2})^2}{(m_1+m_2)^2}}
     e^{-\frac{3\sigma_{32}^2[m_3\bm{p}'_{1}+m_3\bm{p}'_{2}-(m_1+m_2)\bm{p}'_{3}]^2} {2(m_1+m_2+m_3)^2}} .    \label{eq:MHj3h}
\end{eqnarray} }%

We change integral variables in Eq.~(\ref{eq:AHj3h}) to be $\bm{Y}= (m_1\bm{x}_1+m_2\bm{x}_2+m_3\bm{x}_3)/(m_1+m_2+m_3)$, $\bm{r}_1= (\bm{x}_1-\bm{x}_2)/\sqrt{2}$ and 
$\bm{r}_2=\sqrt{\frac{2}{3}} (\frac{m_1\bm{x}_1}{m_1+m_2}+\frac{m_2\bm{x}_2}{m_1+m_2}-\bm{x}_3)$, 
and further assume the coordinate joint distribution is coordinate variable factorized, i.e., 
$3^{3/2} f^{(n)}_{h_1h_2h_3}(\bm{Y},\bm{r}_1,\bm{r}_2) = f^{(n)}_{h_1h_2h_3}(\bm{Y}) f^{(n)}_{h_1h_2h_3}(\bm{r}_1) f^{(n)}_{h_1h_2h_3}(\bm{r}_2)$.
Adopting $f^{(n)}_{h_1h_2h_3}(\bm{r}_1) = \frac{1}{(\pi C_1 R_f^2)^{3/2}} e^{-\frac{\bm{r}_1^2}{C_1 R_f^2}}$ 
and $f^{(n)}_{h_1h_2h_3}(\bm{r}_2) = \frac{1}{(\pi C_2 R_f^2)^{3/2}} e^{-\frac{\bm{r}_2^2}{C_2 R_f^2}}$ as in Refs.~\cite{Mrowczynski:2016xqm,Wang:2020zaw}, we have
{\setlength\arraycolsep{0pt}
\begin{eqnarray}
 \mathcal {A}_{H_j} &=& 8^2  \frac{1}{(\pi C_1 R_f^2)^{3/2}} \int d\bm{r}_1 e^{-\frac{\bm{r}_1^2}{C_1 R_f^2}} e^{-\frac{(\bm{r}'_1)^2}{\sigma_{31}^2}}  \nonumber  \\
   && \times         \frac{1}{(\pi C_2 R_f^2)^{3/2}} \int d\bm{r}_2 e^{-\frac{\bm{r}_2^2}{C_2 R_f^2}} e^{-\frac{(\bm{r}'_2)^2}{\sigma_{32}^2}}.   \label{eq:AHj3h-r}  
\end{eqnarray} }%
Comparing relations of $\bm{r}_1$, $\bm{r}_2$ with $\bm{x}_1$, $\bm{x}_2$, $\bm{x}_3$ to that of $\bm{r}$ with $\bm{x}_1$, $\bm{x}_2$ in Sec.~\ref{2hcoHj}, 
we see that $C_1$ is 2 and $C_2$ is $8/3$~\cite{Mrowczynski:2016xqm,Wang:2020zaw}. 
Considering the coordinate Lorentz transformation and integrating from the relative coordinate variables in Eq.~(\ref{eq:AHj3h-r}), we obtain 
{\setlength\arraycolsep{0pt}
\begin{eqnarray}
 \mathcal {A}_{H_j} &=&  \frac{8^2 \sigma_{31}^3} {  (C_1 R_f^2+\sigma_{31}^2) \sqrt{C_1 (R_f/\gamma)^2+\sigma_{31}^2} }  \nonumber  \\
  && \times \frac{\sigma_{32}^3}{ (C_2 R_f^2+\sigma_{32}^2) \sqrt{C_2 (R_f/\gamma)^2+\sigma_{32}^2}} .   \label{eq:AHj3h-final}  
\end{eqnarray} }%

Approximating the gaussian form of the momentum-dependent kernel function to be $\delta$ function form 
and integrating $\bm{p}_1$, $\bm{p}_2$ and $\bm{p}_3$ from Eq.~(\ref{eq:MHj3h}), we can obtain
{\setlength\arraycolsep{0.2pt}
\begin{eqnarray}
 && \mathcal {M}_{H_j}(\bm{p}) =  \left( \frac{\pi }{\sqrt{3}\sigma_{31}\sigma_{32}}  \right)^3   \gamma^2 \times  \nonumber   \\
  && ~~ f^{(n)}_{h_1h_2h_3}(\frac{m_1\bm{p}}{m_1+m_2+m_3},\frac{m_2\bm{p}}{m_1+m_2+m_3},\frac{m_3\bm{p}}{m_1+m_2+m_3}). ~~~~~ \label{eq:MHj3h-pfin}
\end{eqnarray} }%

Substituting Eqs.~(\ref{eq:AHj3h-final}) and (\ref{eq:MHj3h-pfin}) into Eq.~(\ref{eq:fHj3h}) and ignoring correlations among $h_1$, $h_2$ and $h_3$ hadrons, we have
{\setlength\arraycolsep{0.2pt}
\begin{eqnarray}
&& f_{H_j}(\bm{p})  = \frac{8^2 \pi^3 g_{H_j} \gamma^2}{3^{3/2}(C_1 R_f^2+\sigma_{31}^2) \sqrt{C_1 (R_f/\gamma)^2+\sigma_{31}^2}}     \nonumber   \\
 &&~~~ \times   \frac{1}{ (C_2 R_f^2+\sigma_{32}^2) \sqrt{C_2 (R_f/\gamma)^2+\sigma_{32}^2}}   f_{h_1}(\frac{m_1\bm{p}}{m_1+m_2+m_3})       \nonumber   \\
 &&~~~ \times   f_{h_2}(\frac{m_2\bm{p}}{m_1+m_2+m_3})    f_{h_3}(\frac{m_3\bm{p}}{m_1+m_2+m_3}) .     \label{eq:fHj3h-approx}
\end{eqnarray} }%
Finally, we have the Lorentz invariant momentum distribution
{\setlength\arraycolsep{0.2pt}
\begin{eqnarray}
 && f_{H_j}^{(\text{inv})}(p_{T},y) = \frac{8^2 \pi^3 g_{H_j} }{3^{3/2}(C_1 R_f^2+\sigma_{31}^2) \sqrt{C_1 (R_f/\gamma)^2+\sigma_{31}^2}}     \nonumber   \\
 && ~~~~~~ \times  \frac{1}{ (C_2 R_f^2+\sigma_{32}^2) \sqrt{C_2 (R_f/\gamma)^2+\sigma_{32}^2}}   \frac{m_{H_j}}{m_1m_2m_3}   \nonumber   \\
  &&  ~~~~~~ \times        f_{h_1}^{(\text{inv})}(\frac{m_1p_{T}}{m_1+m_2+m_3},y)  f_{h_2}^{(\text{inv})}(\frac{m_2p_{T}}{m_1+m_2+m_3},y)      \nonumber   \\
  &&  ~~~~~~ \times      f_{h_3}^{(\text{inv})}(\frac{m_3p_{T}}{m_1+m_2+m_3},y) .     \label{eq:pt-Hj3h}
\end{eqnarray} }%

\subsection{Formalism of four-body coalescence}       \label{4hcoHj}

For $H_j$ formed via the coalescence of four hadronic bodies $h_1$, $h_2$, $h_3$ and $h_4$, the three-dimensional momentum distribution is
{\setlength\arraycolsep{0pt}
\begin{eqnarray}
&& f_{H_j}(\bm{p})= N_{h_1h_2h_3h_4} \int d\bm{x}_1d\bm{x}_2d\bm{x}_3d\bm{x}_4 d\bm{p}_1 d\bm{p}_2d\bm{p}_3d\bm{p}_4    \nonumber  \\
&&~~~~~~~~~~~~~~\times f^{(n)}_{h_1h_2h_3h_4}(\bm{x}_1,\bm{x}_2,\bm{x}_3,\bm{x}_4;\bm{p}_1,\bm{p}_2,\bm{p}_3,\bm{p}_4)  \nonumber  \\
&&~~~~~~~~~~~~~~\times \mathcal {R}_{H_j}(\bm{x}_1,\bm{x}_2,\bm{x}_3,\bm{x}_4;\bm{p}_1,\bm{p}_2,\bm{p}_3,\bm{p}_4,\bm{p}).     \label{eq:fHj4general1}               
\end{eqnarray} }%
The number of all possible $h_1h_2h_3h_4$-clusters $N_{h_1h_2h_3h_4}$ equals to $N_{h_1}N_{h_2}N_{h_3}N_{h_4},~N_{h_1}(N_{h_1}-1)N_{h_3}N_{h_4},~N_{h_1}(N_{h_1}-1)N_{h_3}(N_{h_3}-1)$ for $h_1 \neq h_2 \neq h_3 \neq h_4$, $h_1 = h_2 \neq h_3 \neq h_4$, $h_1 = h_2 \neq h_3 = h_4$, respectively.
$f^{(n)}_{h_1h_2h_3h_4}$ is the normalized four-hadron joint coordinate-momentum distribution.

We rewrite the kernel function as
{\setlength\arraycolsep{0pt}
\begin{eqnarray}
&&  \mathcal {R}_{H_j}(\bm{x}_1,\bm{x}_2,\bm{x}_3,\bm{x}_4;\bm{p}_1,\bm{p}_2,\bm{p}_3,\bm{p}_4,\bm{p}) = g_{H_j}  \nonumber  \\
&& ~~~~ \times \mathcal {R}_{H_j}^{(x,p)}(\bm{x}_1,\bm{x}_2,\bm{x}_3,\bm{x}_4;\bm{p}_1,\bm{p}_2,\bm{p}_3,\bm{p}_4) \delta(\displaystyle{\sum^4_{i=1}} \bm{p}_i-\bm{p}),      \label{eq:RHj4fac}  
\end{eqnarray} }%
where the spin degeneracy factor $g_{H_j} = (2J_{H_j}+1) /[\prod \limits_{i=1}^4(2J_{h_i}+1)]$, and
\begin{widetext}
{\setlength\arraycolsep{0pt}
\begin{eqnarray}
  \mathcal {R}^{(x,p)}_{H_j}(\bm{x}_1,\bm{x}_2,\bm{x}_3,\bm{x}_4;\bm{p}_1,\bm{p}_2,\bm{p}_3,\bm{p}_4) &=& 
8^3 e^{-\frac{(\bm{x}'_1-\bm{x}'_2)^2}{2\sigma_{41}^2}}
      e^{-\frac{2(\frac{m_1\bm{x}'_1}{m_1+m_2}+\frac{m_2\bm{x}'_2}{m_1+m_2}-\bm{x}'_3)^2}{3\sigma_{42}^2}}
      e^{-\frac{3(\frac{m_1\bm{x}'_1}{m_1+m_2+m_3}+\frac{m_2\bm{x}'_2}{m_1+m_2+m_3}+\frac{m_3\bm{x}'_3}{m_1+m_2+m_3}-\bm{x}'_4)^2}{4\sigma_{43}^2}}  \nonumber \\
&&\times e^{-\frac{2\sigma_{41}^2(m_2\bm{p}'_{1}-m_1\bm{p}'_{2})^2}{(m_1+m_2)^2}}
     e^{-\frac{3\sigma_{42}^2[m_3\bm{p}'_{1}+m_3\bm{p}'_{2}-(m_1+m_2)\bm{p}'_{3}]^2} {2(m_1+m_2+m_3)^2}}
    e^{-\frac{4\sigma_{43}^2[m_4\bm{p}'_{1}+m_4\bm{p}'_{2}+m_4\bm{p}'_{3}-(m_1+m_2+m_3)\bm{p}'_{4}]^2} {3(m_1+m_2+m_3+m_4)^2}}.
\end{eqnarray} }%
Here, the width parameter
$\sigma_{41}=\sqrt{\frac{4(m_1+m_2)m_3m_4(m_1+m_2+m_3+m_4)} {3[m_1m_2m_3(m_1+m_2+m_3)+m_2m_3m_4(m_2+m_3+m_4)+m_3m_4m_1(m_3+m_4+m_1)+m_4m_1m_2(m_4+m_1+m_2)]}} RMS_{H_j}$, \\
$\sigma_{42}=\sqrt{\frac{16(m_1+m_2+m_3)m_1m_2m_4(m_1+m_2+m_3+m_4)} {9(m_1+m_2)[m_1m_2m_3(m_1+m_2+m_3)+m_2m_3m_4(m_2+m_3+m_4)+m_3m_4m_1(m_3+m_4+m_1)+m_4m_1m_2(m_4+m_1+m_2)]}} RMS_{H_j}$, and
$\sigma_{43}=\sqrt{\frac{2m_1m_2m_3(m_1+m_2+m_3+m_4)^2} {(m_1+m_2+m_3)[m_1m_2m_3(m_1+m_2+m_3)+m_2m_3m_4(m_2+m_3+m_4)+m_3m_4m_1(m_3+m_4+m_1)+m_4m_1m_2(m_4+m_1+m_2)]}} RMS_{H_j}$.

Assuming that the normalized joint distribution is coordinate and momentum factorized, we have
{\setlength\arraycolsep{0.2pt}
\begin{eqnarray}
 f_{H_j}(\bm{p}) 
= N_{h_1h_2h_3h_4} g_{H_j} \mathcal {A}_{H_j}   \mathcal {M}_{H_j}(\bm{p}).     \label{eq:fHj4}  
\end{eqnarray} }%
In Eq.~(\ref{eq:fHj4}), we use $\mathcal {A}_{H_j}$ to denote the coordinate integral part  as
{\setlength\arraycolsep{0pt}
\begin{eqnarray}
\mathcal {A}_{H_j} =  8^3  \int d\bm{x}_1d\bm{x}_2d\bm{x}_3d\bm{x}_4 f^{(n)}_{h_1h_2h_3h_4}(\bm{x}_1,\bm{x}_2,\bm{x}_3,\bm{x}_4)   
      e^{-\frac{(\bm{x}'_1-\bm{x}'_2)^2}{2\sigma_{41}^2}}
      e^{-\frac{2(\frac{m_1\bm{x}'_1}{m_1+m_2}+\frac{m_2\bm{x}'_2}{m_1+m_2}-\bm{x}'_3)^2}{3\sigma_{42}^2}}
      e^{-\frac{3(\frac{m_1\bm{x}'_1}{m_1+m_2+m_3}+\frac{m_2\bm{x}'_2}{m_1+m_2+m_3}+\frac{m_3\bm{x}'_3}{m_1+m_2+m_3}-\bm{x}'_4)^2}{4\sigma_{43}^2}} ,      \label{eq:AHj4}      
\end{eqnarray} }%
and $\mathcal {M}_{H_j}(\bm{p})$ to denote the momentum integral part as
{\setlength\arraycolsep{0pt}
\begin{eqnarray}
   \mathcal {M}_{H_j}(\bm{p}) &=&   \int d\bm{p}_1d\bm{p}_2d\bm{p}_3d\bm{p}_4 f^{(n)}_{h_1h_2h_3h_4}(\bm{p}_1,\bm{p}_2,\bm{p}_3,\bm{p}_4) 
     e^{-\frac{2\sigma_{41}^2(m_2\bm{p}'_{1}-m_1\bm{p}'_{2})^2}{(m_1+m_2)^2}}
     e^{-\frac{3\sigma_{42}^2[m_3\bm{p}'_{1}+m_3\bm{p}'_{2}-(m_1+m_2)\bm{p}'_{3}]^2} {2(m_1+m_2+m_3)^2}}
    e^{-\frac{4\sigma_{43}^2[m_4\bm{p}'_{1}+m_4\bm{p}'_{2}+m_4\bm{p}'_{3}-(m_1+m_2+m_3)\bm{p}'_{4}]^2} {3(m_1+m_2+m_3+m_4)^2}}  \nonumber  \\
&& ~~~~ \times   \delta(\displaystyle{\sum^4_{i=1}} \bm{p}_i-\bm{p}).    \label{eq:MHj4}
\end{eqnarray} }%

We change integral variables in Eq.~(\ref{eq:AHj4}) to be $\bm{Z}= (m_1\bm{x}_1+m_2\bm{x}_2+m_3\bm{x}_3+m_4\bm{x}_4)/(m_1+m_2+m_3+m_4)$,
$\bm{r}_1= (\bm{x}_1-\bm{x}_2)/\sqrt{2}$, $\bm{r}_2=\sqrt{\frac{2}{3}} (\frac{m_1\bm{x}_1}{m_1+m_2}+\frac{m_2\bm{x}_2}{m_1+m_2}-\bm{x}_3)$ and
$\bm{r}_3=\sqrt{\frac{3}{4}} (\frac{m_1\bm{x}_1}{m_1+m_2+m_3}+\frac{m_2\bm{x}_2}{m_1+m_2+m_3}+\frac{m_3\bm{x}_3}{m_1+m_2+m_3}-\bm{x}_4)$, 
and assume $f^{(n)}_{h_1h_2h_3h_4}(\bm{Z},\bm{r}_1,\bm{r}_2,\bm{r}_3) = f^{(n)}_{h_1h_2h_3h_4}(\bm{Z}) f^{(n)}_{h_1h_2h_3h_4}(\bm{r}_1) f^{(n)}_{h_1h_2h_3h_4}(\bm{r}_2)$ $ f^{(n)}_{h_1h_2h_3h_4}(\bm{r}_3)$.
Adopting $f^{(n)}_{h_1h_2h_3h_4}(\bm{r}_1) = \frac{1}{(\pi C_1 R_f^2)^{3/2}} e^{-\frac{\bm{r}_1^2}{C_1 R_f^2}}$,
$f^{(n)}_{h_1h_2h_3h_4}(\bm{r}_2) = \frac{1}{(\pi C_2 R_f^2)^{3/2}} e^{-\frac{\bm{r}_2^2}{C_2 R_f^2}}$ and $f^{(n)}_{h_1h_2h_3h_4}(\bm{r}_3) = \frac{1}{(\pi C_3 R_f^2)^{3/2}} e^{-\frac{\bm{r}_3^2}{C_3 R_f^2}}$, we have
{\setlength\arraycolsep{0pt}
\begin{eqnarray}
 \mathcal {A}_{H_j} =  8^3 \frac{1}{(\pi C_1 R_f^2)^{3/2}} \int d\bm{r}_1 e^{-\frac{\bm{r}_1^2}{C_1 R_f^2}} e^{-\frac{(\bm{r}'_1)^2}{\sigma_{41}^2}} 
    \frac{1}{(\pi C_2 R_f^2)^{3/2}} \int d\bm{r}_2 e^{-\frac{\bm{r}_2^2}{C_2 R_f^2}} e^{-\frac{(\bm{r}'_2)^2}{\sigma_{42}^2}}  
    \frac{1}{(\pi C_3 R_f^2)^{3/2}} \int d\bm{r}_3 e^{-\frac{\bm{r}_3^2}{C_3 R_f^2}} e^{-\frac{(\bm{r}'_3)^2}{\sigma_{43}^2}}.      \label{eq:AHj4-r}  
\end{eqnarray} }%
$C_1$, $C_2$, $C_3$ are equal to $2$, $8/3$ and $3$, respectively ~\cite{Mrowczynski:2016xqm,Wang:2020zaw}. 
After the Lorentz transformation and integrating the relative coordinate variables from Eq.~(\ref{eq:AHj4-r}), we obtain 
{\setlength\arraycolsep{0pt}
\begin{eqnarray}
  \mathcal {A}_{H_j} =  \frac{8^3 \sigma_{41}^3} {(C_1 R_f^2+\sigma_{41}^2) \sqrt{C_1 (R_f/\gamma)^2+\sigma_{41}^2} }  
   \frac{\sigma_{42}^3}{(C_2 R_f^2+\sigma_{42}^2) \sqrt{C_2 (R_f/\gamma)^2+\sigma_{42}^2} }  
   \frac{\sigma_{43}^3}{(C_3 R_f^2+\sigma_{43}^2) \sqrt{C_3 (R_f/\gamma)^2+\sigma_{43}^2}} .   \label{eq:AHj4-fin}  
\end{eqnarray} }%

Approximating the gaussian form of the momentum-dependent kernel function to be $\delta$ function form and after integrating $\bm{p}_1$, $\bm{p}_2$, $\bm{p}_3$ and $\bm{p}_4$ in Eq.~(\ref{eq:MHj4}),  we can obtain
{\setlength\arraycolsep{0.2pt}
\begin{eqnarray}
\mathcal {M}_{H_j}(\bm{p}) = (\frac{\pi^{3/2}\gamma}{\sqrt{4}\sigma_{41}\sigma_{42}\sigma_{43}})^3  f^{(n)}_{h_1h_2h_3h_4}(\frac{m_1\bm{p}}{m_1+m_2+m_3+m_4},\frac{m_2\bm{p}}{m_1+m_2+m_3+m_4},\frac{m_3\bm{p}}{m_1+m_2+m_3+m_4},\frac{m_4\bm{p}}{m_1+m_2+m_3+m_4}).~~~~~~    \label{eq:MHj4-fin}  
\end{eqnarray} }%

Substituting Eqs.~(\ref{eq:AHj4-fin}) and (\ref{eq:MHj4-fin}) into Eq.~(\ref{eq:fHj4}) and ignoring correlations among $h_1$, $h_2$, $h_3$ and $h_4$ hadrons, we have
{\setlength\arraycolsep{0.2pt}
\begin{eqnarray}
 f_{H_j}(\bm{p}) &=&\frac{8^3 (\pi^{3/2})^3  g_{H_j}  \gamma^3}
 {4^{3/2}(C_1 R_f^2+\sigma_{41}^2) \sqrt{C_1 (R_f/\gamma)^2+\sigma_{41}^2} }      \frac{1}{(C_2 R_f^2+\sigma_{42}^2) \sqrt{C_2 (R_f/\gamma)^2+\sigma_{42}^2} }  
   \frac{1}{ (C_3 R_f^2+\sigma_{43}^2) \sqrt{C_3 (R_f/\gamma)^2+\sigma_{43}^2}}    \nonumber  \\
&& \times   f_{h_1}(\frac{m_1\bm{p}}{m_1+m_2+m_3+m_4})  f_{h_2}(\frac{m_2\bm{p}}{m_1+m_2+m_3+m_4}) 
                  f_{h_3}(\frac{m_3\bm{p}}{m_1+m_2+m_3+m_4})     f_{h_4}(\frac{m_4\bm{p}}{m_1+m_2+m_3+m_4})  .  \label{eq:fHj4-approx}
\end{eqnarray} }%
We finally have the Lorentz invariant momentum distribution
{\setlength\arraycolsep{0.2pt}
\begin{eqnarray}
   f_{H_j}^{(\text{inv})}(p_{T},y) &=& \frac{8^3 (\pi^{3/2})^3  g_{H_j}}
 {4^{3/2}(C_1 R_f^2+\sigma_{41}^2) \sqrt{C_1 (R_f/\gamma)^2+\sigma_{41}^2} }      \frac{1}{(C_2 R_f^2+\sigma_{42}^2) \sqrt{C_2 (R_f/\gamma)^2+\sigma_{42}^2} }  
   \frac{1}{ (C_3 R_f^2+\sigma_{43}^2) \sqrt{C_3 (R_f/\gamma)^2+\sigma_{43}^2}}   \nonumber  \\
&& \times    \frac{m_{H_j}}{m_1m_2m_3m_4}   f_{h_1}^{(\text{inv})}(\frac{m_1p_T}{m_1+m_2+m_3+m_4},y)  f_{h_2}^{(\text{inv})}(\frac{m_2p_T}{m_1+m_2+m_3+m_4},y) 
                  f_{h_3}^{(\text{inv})}(\frac{m_3p_T}{m_1+m_2+m_3+m_4},y)     \nonumber  \\
&& \times     f_{h_4}^{(\text{inv})}(\frac{m_4p_T}{m_1+m_2+m_3+m_4},y) .  \label{eq:pt-Hj4h}
\end{eqnarray} }%

\end{widetext}

\subsection{General formalism of $N$-body coalescence}

From the formula derivations of two-body, three-body, and four-body coalescences, we can get a general formalism of $N$-body ($N=2,~3,~4,\cdot\cdot\cdot$) coalescence.
For the three-dimensional momentum distribution of $H_j$ formed via the coalescence of $N$ hadronic bodies $h_1$, $h_2$, $\cdot\cdot\cdot$ and $h_N$, it is 
{\setlength\arraycolsep{0pt}
\begin{eqnarray}
&& f_{H_j}(\bm{p})= \int \prod \limits_{i=1}^{N}d\bm{x}_i d\bm{p}_i  f_{h_1h_2\cdot\cdot\cdot h_N}(\bm{x}_1,\bm{x}_2,\cdot\cdot\cdot,\bm{x}_N;\bm{p}_1,\bm{p}_2,\cdot\cdot\cdot,\bm{p}_N)  \nonumber  \\
&&~~~~~~~~~~~~~~~~~~~~\times \mathcal {R}_{H_j}(\bm{x}_1,\bm{x}_2,\cdot\cdot\cdot,\bm{x}_N;\bm{p}_1,\bm{p}_2,\cdot\cdot\cdot,\bm{p}_N,\bm{p}).     \label{eq:fHjngeneral1}               
\end{eqnarray} }%
$f_{h_1h_2\cdot\cdot\cdot h_N}$ is the $N$-hadron joint coordinate-momentum distribution and $\mathcal {R}_{H_j}$ is the kernel function.

Adopting a spherical harmonic oscillator wave function of $H_j$ and the coordinate-momentum factorization assumption of $f_{h_1h_2\cdot\cdot\cdot h_N}$, we can get
{\setlength\arraycolsep{0.2pt}
\begin{eqnarray}
 f_{H_j}(\bm{p}) 
= N_{h_1h_2\cdot\cdot\cdot h_N} g_{H_j} \mathcal {A}_{H_j}   \mathcal {M}_{H_j}(\bm{p}).     \label{eq:fHjn}  
\end{eqnarray} }%
$N_{h_1h_2\cdot\cdot\cdot h_N}$ is the number of all possible $(h_1h_2\cdot\cdot\cdot h_N)$-clusters in the considered hadronic system.
The spin degeneracy factor $g_{H_j} = (2J_{H_j}+1) /[\prod \limits_{i=1}^N(2J_{h_i}+1)]$, where $J_{H_j}$ and $J_{h_i}$ are the spin of the $H_j$ and that of the $h_i$.
$\mathcal {A}_{H_j}$ stands for the probability of one $(h_1h_2\cdot\cdot\cdot h_N)$-cluster satisfying the coordinate requirement to recombine into $H_j$,
and $\mathcal {M}_{H_j}(\bm{p})$ stands for the probability density of one $(h_1h_2\cdot\cdot\cdot h_N)$-cluster satisfying the momentum requirement to recombine into $H_j$ with momentum $\bm{p}$.
They are
{\setlength\arraycolsep{0pt}
\begin{eqnarray}
  \mathcal {A}_{H_j} =  8^{N-1} \prod \limits_{k=1}^{N-1} \frac{ \sigma_{Nk}^3} {(C_k R_f^2+\sigma_{Nk}^2) \sqrt{C_k (R_f/\gamma)^2+\sigma_{Nk}^2} },   \label{eq:AHjn-fin}  
\end{eqnarray} }%
and
{\setlength\arraycolsep{0.2pt}
\begin{eqnarray}
\mathcal {M}_{H_j}(\bm{p}) &=& \frac{\pi^{3(N-1)/2}\gamma^{N-1}}{N^{3/2} \prod \limits_{k=1}^{N-1}\sigma_{Nk}^3}   f^{(n)}_{h_1h_2\cdot\cdot\cdot h_N}(\frac{m_1\bm{p}}{m_1+m_2+...+m_N},    \nonumber \\
&& \frac{m_2\bm{p}}{m_1+m_2+...+m_N},\cdot\cdot\cdot,\frac{m_N\bm{p}}{m_1+m_2+...+m_N}).~~~~~~    \label{eq:MHjn-fin}  
\end{eqnarray} }%
Here, $C_k$ is the coefficient in the relative coordinate distribution and it has certain values as stated in Secs.~\ref{2hcoHj}, \ref{3hcoHj}, and \ref{4hcoHj}.
The gaussian width parameter $\sigma_{Nk}$ comes from the kernel function. $R_f$ is the effective radius of the hadronic system at the $H_j$ freeze-out.
$f^{(n)}_{h_1h_2\cdot\cdot\cdot h_N}$ is the normalized $N$-hadron momentum distribution.
$m_i~(i=1,~2,\cdot\cdot\cdot,~N)$ is the mass of the hadron $h_i$.

Ignoring correlations in momentum distributions of $h_i~(i=1,~2,\cdot\cdot\cdot,~N)$ hadrons, the invariant momentum distribution of $H_j$ formed via $N$-body coalescence can be expressed as
{\setlength\arraycolsep{0.2pt}
\begin{eqnarray}
   f_{H_j}^{(\text{inv})}(p_{T},y)& =& \frac{8^{N-1} \pi^{3(N-1)/2}  g_{H_j}m_{H_j} }   {N^{3/2} \prod \limits_{k=1}^{N-1} \left(C_k R_f^2+\sigma_{Nk}^2\right) \sqrt{C_k (R_f/\gamma)^2+\sigma_{Nk}^2} }         \nonumber  \\
&& \times    \prod \limits_{i=1}^{N} \frac{1}{m_i} f_{h_i}^{(\text{inv})}(\frac{m_ip_T}{m_1+m_2+...+m_N},y)  .  \label{eq:pt-Hjnh}
\end{eqnarray} }%
Eq.~(\ref{eq:pt-Hjnh}) gives the relationship of light (hyper-)nuclei with primordial hadronic bodies in momentum space in the laboratory frame. 
It can be directly used to calculate the rapidity and $p_T$ distributions of light (hyper-)nuclei formed via different coalescence channels as long as the primordial momentum distributions of nucleons and hyperons are given.
In the following section, we will show its applications in heavy ion collisions at midrapidity ($y=0$), where is usually covered by the experimental detectors.

\section{Results and discussions}   \label{results}

In this section, we apply the coalescence model to central Au-Au collisions at $\sqrt{s_{NN}}=3$ GeV to study the productions of $^3_{\Lambda}$H, $^4_{\Lambda}$H and $^4_{\Lambda}$He in low- and intermediate-$p_T$ regions at midrapidity.
We first introduce the necessary model inputs, i.e., $p_T$ spectra of nucleons and $\Lambda$ hyperons. 
Then we present $p_T$-dependent and $p_T$-integrated contributions from different coalescence channels for $^3_{\Lambda}$H, $^4_{\Lambda}$H and $^4_{\Lambda}$He.
Finally we investigate the influences of $^2_{\Lambda}n$ and $^3_{\Lambda}n$, if they exist, on the productions of $^3_{\Lambda}$H, $^4_{\Lambda}$H and $^4_{\Lambda}$He. 
We propose the production asymmetry between $^4_{\Lambda}$H and $^4_{\Lambda}$He as an effective probe to further help judge the existences of $^2_{\Lambda}n$ and $^3_{\Lambda}n$.

\subsection{$p_T$ spectra of nucleons and $\Lambda$ hyperons}

The invariant $p_T$ distributions of primordial protons $f_{p,\text{pri}}^{(\text{inv})}(p_{T})$, neutrons $f_{n,\text{pri}}^{(\text{inv})}(p_{T})$ and $\Lambda$ hyperons $f_{\Lambda,\text{pri}}^{(\text{inv})}(p_{T})$ are necessary inputs for computing $p_T$ distributions of $\Lambda$-hypernuclei in the coalescence model.
We here use the blast-wave model~\cite{Schnedermann:1993ws} to get these invariant $p_T$ distribution functions by fitting the measured data of primordial protons and $\Lambda$'s in Refs.~\cite{STAR:2023uxk,STAR:2024znc}.

\begin{figure}[htbp]
\centering
 \includegraphics[width=0.95\linewidth]{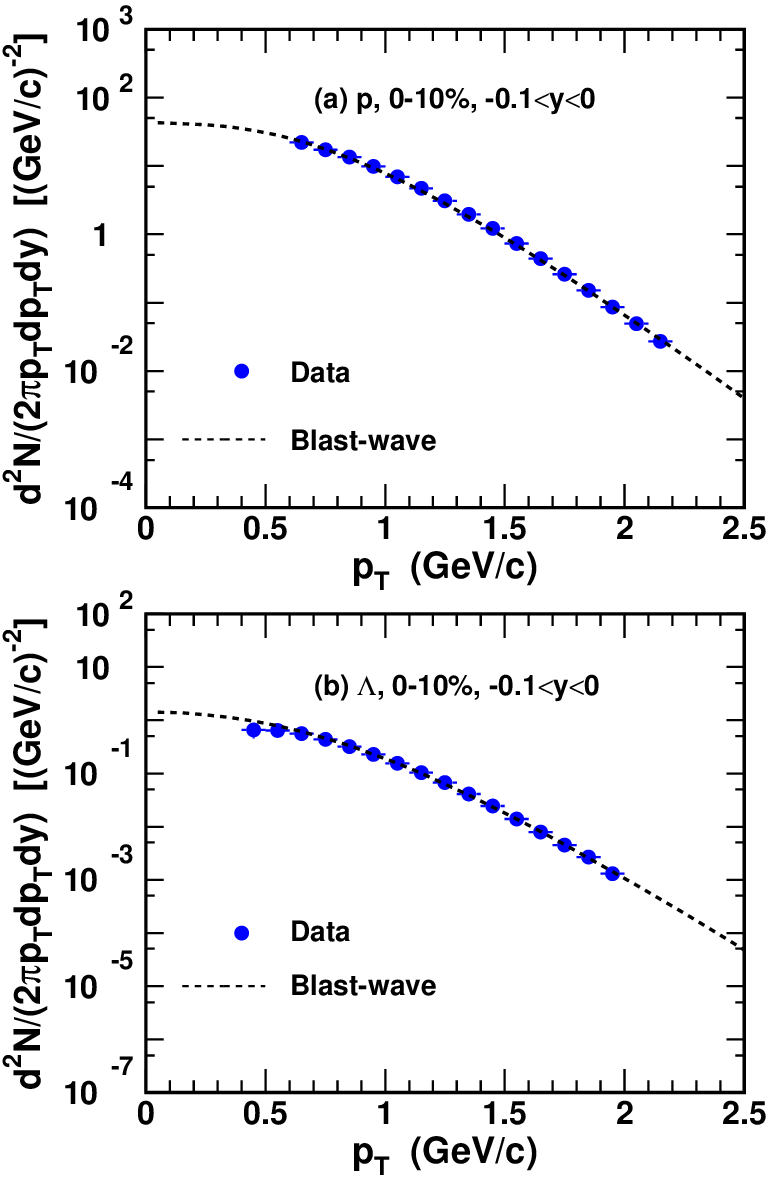}\\
 \caption{Invariant $p_T$ spectra of primordial (a) $p$ and (b) $\Lambda$ measured experimentally at midrapidity in central Au-Au collisions at $\sqrt{s_{NN}}=3$ GeV. Filled circles with error bars are experimental data~\cite{STAR:2023uxk,STAR:2024znc}. Dashed lines are the results of the blast-wave model. }
 \label{fig:pLambdapT}
\end{figure}

Fig.~\ref{fig:pLambdapT} shows the invariant $p_T$ spectra of primordial protons and $\Lambda$ hyperons measured experimentally at the rapidity interval $-0.1<y<0$ in the $0-10$\% centrality in Au-Au collisions at $\sqrt{s_{NN}}=3$ GeV.
Filled circles with error bars are experimental data~\cite{STAR:2023uxk,STAR:2024znc}. 
Dashed lines are the results of the blast-wave model.
Note that protons and $\Lambda$'s shown in Fig.~\ref{fig:pLambdapT} are these primordial ones measured at the experiment.
In fact the total primordial protons and $\Lambda$'s in $f_{p,\text{pri}}^{(\text{inv})}(p_{T})$ and $f_{\Lambda,\text{pri}}^{(\text{inv})}(p_{T})$ should include those consumed ones entering into light nuclei and hypernuclei.
The consumed protons and $\Lambda$'s take about 20\% and less than 1\%, respectively~\cite{STAR:2023uxk,STAR:2024znc,STAR:2021orx}.
So dividing protons in Fig.~\ref{fig:pLambdapT} (a) by 80\%, we get $f_{p,\text{pri}}^{(\text{inv})}(p_{T},y)$.
Due to the negligible consumed $\Lambda$'s, the dashed line in Fig.~\ref{fig:pLambdapT} (b) is just $f_{\Lambda,\text{pri}}^{(\text{inv})}(p_{T})$.

For the neutron, we assume the same normalized $p_T$ distribution as that of the proton.
For absolute yield density of the neutron, it is generally not equal to that of the proton due to the prominent influence of net nucleons from the colliding Au nuclei.
We use $Z_{np}$ to denote the extent of the yield density asymmetry of the neutron and the proton and take their relation as $\frac{dN_n}{dy}=\frac{dN_p}{dy}\times Z_{np}$.
Note that $Z_{np}=1$ corresponds to the complete isospin equilibration and $Z_{np}=1.49$ to isospin asymmetry in the whole Au nucleus.
We here use $Z_{np}=1.34$, which has been fixed by the experimental data of the yield ratio $t/^3$He~\cite{STAR:2023uxk}.
The other parameter in our model is the effective radius of the hadronic system $R_f$.
It has been determined to be 3.27 fm by reproducing the data of the $d$ yield rapidity density~\cite{STAR:2023uxk}, which is located at the area $3.0\sim3.4$ fm extracted from measurements of the $p-\Lambda$ and $d-\Lambda$ correlations~\cite{Hu:2023iib}.
Both of the parameters $Z_{np}$ and $R_f$ have been determined by the light nucleus production in Ref.~\cite{Wang:2022hja}.
There are no additional free parameters in the current study for hypernuclei.
Therefore, our results of hypernuclei are more potent for further tests of the coalescence mechanism in describing the productions of nuclei with strangeness flavor quantum number.

\subsection{Results of $^3_{\Lambda}$H and $^4_{\Lambda}$H}  \label{IIIB}

\begin{figure}[htbp]
\centering
 \includegraphics[width=0.95\linewidth]{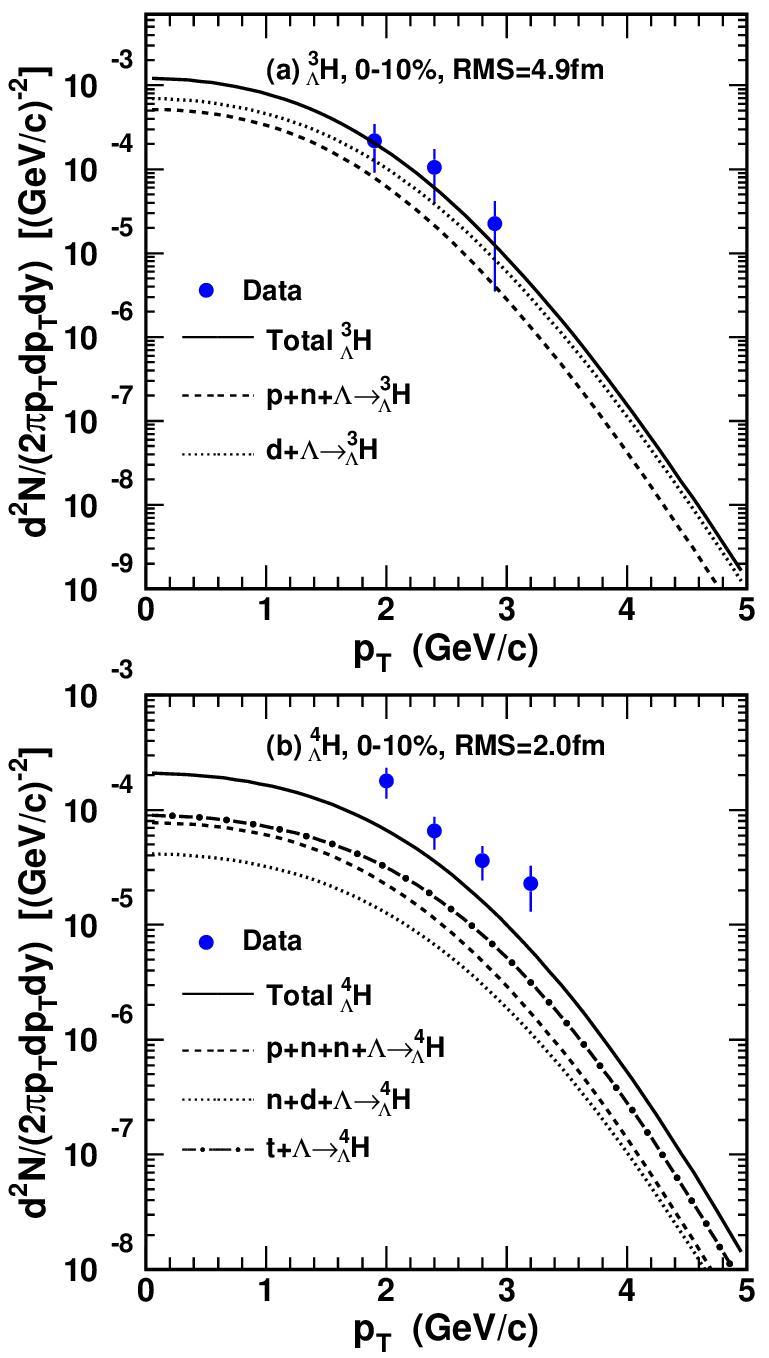}\\
 \caption{Invariant $p_T$ spectra of (a) $^3_{\Lambda}$H and (b) $^4_{\Lambda}$H at midrapidity in central Au-Au collisions at $\sqrt{s_{NN}}=3$ GeV. Filled circles with error bars are experimental data~\cite{STAR:2021orx}, and different lines are the theoretical results.}
 \label{fig:hyt34pT-NonLNonnL}
\end{figure}

We first study the $^3_{\Lambda}$H production.
As its binding energy is rather small compared to light nuclei or heavier hypernuclei~\cite{STAR:2019wjm,ALICE:2022sco,Juric:1973zq}, the $^3_{\Lambda}$H should have a relatively loosely-bound structure and a relatively large size.
It would be easily destroyed after its production and is likely to freeze out at the latest stage, i.e., after the formations of all the other light (hyper-)nuclei~\cite{Zhang:2018euf}.
So nucleons and $\Lambda$'s taking part in $^3_{\Lambda}$H production are approximately those primordial ones measured experimentally rather than the total primordial ones.
Here, two different coalescence channels $p+n+\Lambda$ and $d+\Lambda$ contribute to the $^3_{\Lambda}$H production.
In our calculations the root-mean-square radius of $^3_{\Lambda}$H is $4.9$ fm~\cite{Nemura:1999qp}.
Fig.~\ref{fig:hyt34pT-NonLNonnL} (a) shows the invariant $p_T$ distributions of $^3_{\Lambda}$H in the rapidity interval $-0.1<y<0$ in $0-10$\% centrality in Au-Au collisions at $\sqrt{s_{NN}}=3$ GeV.
Filled circles with error bars are experimental data~\cite{STAR:2021orx}. 
The dashed line denotes the contribution from the coalescence channel $p+n+\Lambda \rightarrow ^3_{\Lambda}$H, and the dotted line from the channel $d+\Lambda \rightarrow ^3_{\Lambda}$H.
The solid line represents the total result from these two coalescence channels, which can reproduce the experimental data within the large error bars.

We then compute the invariant $p_T$ distributions of $^4_{\Lambda}$H in the rapidity interval $-0.1<y<0$ in $0-10$\% centrality in Au-Au collisions at $\sqrt{s_{NN}}=3$ GeV.
The root-mean-square radius of $^4_{\Lambda}$H is $2.0$ fm~\cite{Nemura:1999qp} in our calculations here, a typical value for light (hyper-)nuclei with several MeV binding energies.
The influences of different values for the root-mean-square radius will be discussed later.
Here three different coalescence channels $p+n+n+\Lambda$, $n+d+\Lambda$, and $t+\Lambda$ to the $^4_{\Lambda}$H production are considered.
Filled circles with error bars in Fig.~\ref{fig:hyt34pT-NonLNonnL} (b) are experimental data~\cite{STAR:2021orx}. 
The dashed, dotted, and dashed-dotted lines denote the contributions from the coalescence channels $p+n+n+\Lambda \rightarrow ^4_{\Lambda}$H, $n+d+\Lambda \rightarrow ^4_{\Lambda}$H, and $t+\Lambda \rightarrow ^4_{\Lambda}$H, respectively.
The solid line represents the total result of $^4_{\Lambda}$H from the above three coalescence channels, which shows a serious underestimation of the experimental data.

\begin{table*}[!htb]
\begin{center}
\caption{Yield rapidity densities $dN/dy$ and averaged transverse momenta $\langle p_T\rangle$ of $^3_{\Lambda}$H and $^4_{\Lambda}$H at midrapidity in central Au-Au collisions at $\sqrt{s_{NN}}=3$ GeV. Experimental data in the third and fifth columns are from Refs.~\cite{STAR:2021orx,Ji:2023pog}. Theoretical results are in the fourth and sixth columns.} 
\label{table:dNdyavept-hyt34}
\begin{tabular}{@{\extracolsep{\fill}}cccccccc@{\extracolsep{\fill}}}
\toprule
     &\multirow{2}{*}{Coal-channel}     &\multicolumn{2}{@{}c@{}}{$dN/dy$}          &$~~~$          &\multicolumn{2}{@{}c@{}}{$\langle p_T\rangle$} \\
\cline{3-4}  \cline{6-7} 
                          &                               &Data                            &Theory                                       &          &Data    &Theory      \\
\hline
\multirow{3}{*}{$^3_{\Lambda}$H}
 &$p+n+\Lambda$                               &$---$                           &$0.323\times10^{-2}$                  &          &$---$   &$1.191$    \\
 &$d+\Lambda$                                   &$---$                           &$0.467\times10^{-2}$                  &          &$---$   &$1.246$    \\
 &Total                &$(1.131\pm0.210\pm0.384)\times10^{-2}$    &$0.790\times10^{-2}$                  &          &$1.254\pm0.087\pm0.163$   &$1.224$    \\
\hline
\multirow{5}{*}{$^4_{\Lambda}$H}
 &$p+n+n+\Lambda$                          &$---$                           &$0.764\times10^{-3}$                  &          &$---$   &$1.475$    \\
 &$n+d+\Lambda$                              &$---$                           &$0.423\times10^{-3}$                  &          &$---$   &$1.514$    \\
 &$t+\Lambda$                                   &$---$                           &$1.001\times10^{-3}$                  &          &$---$   &$1.566$    \\
 &Total              &$(4.954\pm0.434\pm1.014)\times10^{-3}$    &$2.188\times10^{-3}$                  &          &$1.506\pm0.043\pm0.122$   &$1.524$    \\
\botrule
\end{tabular}
\end{center}
\end{table*}

Table \ref{table:dNdyavept-hyt34} shows yield rapidity densities $dN/dy$ and averaged transverse momenta $\langle p_T\rangle$ of $^3_{\Lambda}$H and $^4_{\Lambda}$H at midrapidity in central Au-Au collisions at $\sqrt{s_{NN}}=3$ GeV. 
Experimental data of $dN/dy$ in the third column are obtained with the decay branching ratio 25\% for $^3_{\Lambda}\text{H} \rightarrow ^3_{\Lambda}\text{He}+\pi^-$ and 50\% for $^4_{\Lambda}\text{H} \rightarrow ^4_{\Lambda}\text{He}+\pi^-$ in Ref.~\cite{STAR:2021orx}. 
Experimental data of $\langle p_T\rangle$ in the fifth column are from Ref.~\cite{Ji:2023pog}. 
Theoretical results from different coalescence channels are in the fourth and sixth columns.
For both $^3_{\Lambda}$H and $^4_{\Lambda}$H, $\langle p_T\rangle$ theoretical results from different channels are almost consistent with each other and all agree with the available data.
The theoretical total result of $dN/dy$ of $^3_{\Lambda}$H is about 30\% smaller than the central value of the data but within the error uncertainties.
For $^4_{\Lambda}$H, the total result of $dN/dy$ is about 56\% smaller, and even 38\% smaller than the lower limit of the experimental data.

From Fig.~\ref{fig:hyt34pT-NonLNonnL} and Table \ref{table:dNdyavept-hyt34}, one can see the coalescence model gives slight underestimation of $^3_{\Lambda}$H and  serious underestimation of $^4_{\Lambda}$H.
One reason for such underestimations could be that decay contributions from the excited states of hypernuclei are not included.
Theoretical studies in Ref.~\cite{Botvina:2012zz} showed that these excited hypernuclei dominantly decay into single $\Lambda$ instead of $^3_{\Lambda}$H or $^4_{\Lambda}$H.
Data-driven investigation on such decay effect is currently lacking due to the deficiency of the experimental dataset on the existences and decay properties of excited hypernuclei. 

Another reason for underestimations of $^3_{\Lambda}$H and $^4_{\Lambda}$H may be the omission of some certain coalescence channels.
In these channels some perdue states which have not been affirmed at experiments participate in the coalescence process, such as possible neutron-$\Lambda$ bound states. 
The most possible two of such perdue states are $^2_{\Lambda}n$ and $^3_{\Lambda}n$.
Many theoretical literatures have argued their existences~\cite{Dalitz:1958dh,Richard:2014pwa,Afnan:2015ahc,Schafer:2020rba,Htun:2022vxx,NPLQCD:2012mex,Gal:2014efa}.
Several experiments such as the HypHI~\cite{HypHI:2013sxa,Saito:2015tpa},  the ALICE~\cite{Barile:2014wma}, and the WASA-FRS~\cite{WASA-FRS:2023yro,Escrig:2024fnr} have executed the search.
The HypHI Collaboration has declared a possible signal of $^3_{\Lambda}n$ in the reaction of $^6$Li+$^{12}$C at $2A$ GeV~\cite{HypHI:2013sxa}, but unfortunately whether it exists as a bound state remains an open question till now.
In this work, we provide an alternative way to examine and study these possible bound states, i.e., via reproducing the multiplicities of $^3_{\Lambda}$H and $^4_{\Lambda}$H in relativistic heavy ion collisions.

\subsection{Predictions of $^2_{\Lambda}n$ and $^3_{\Lambda}n$}

\begin{figure}[htbp]
\centering
 \includegraphics[width=0.95\linewidth]{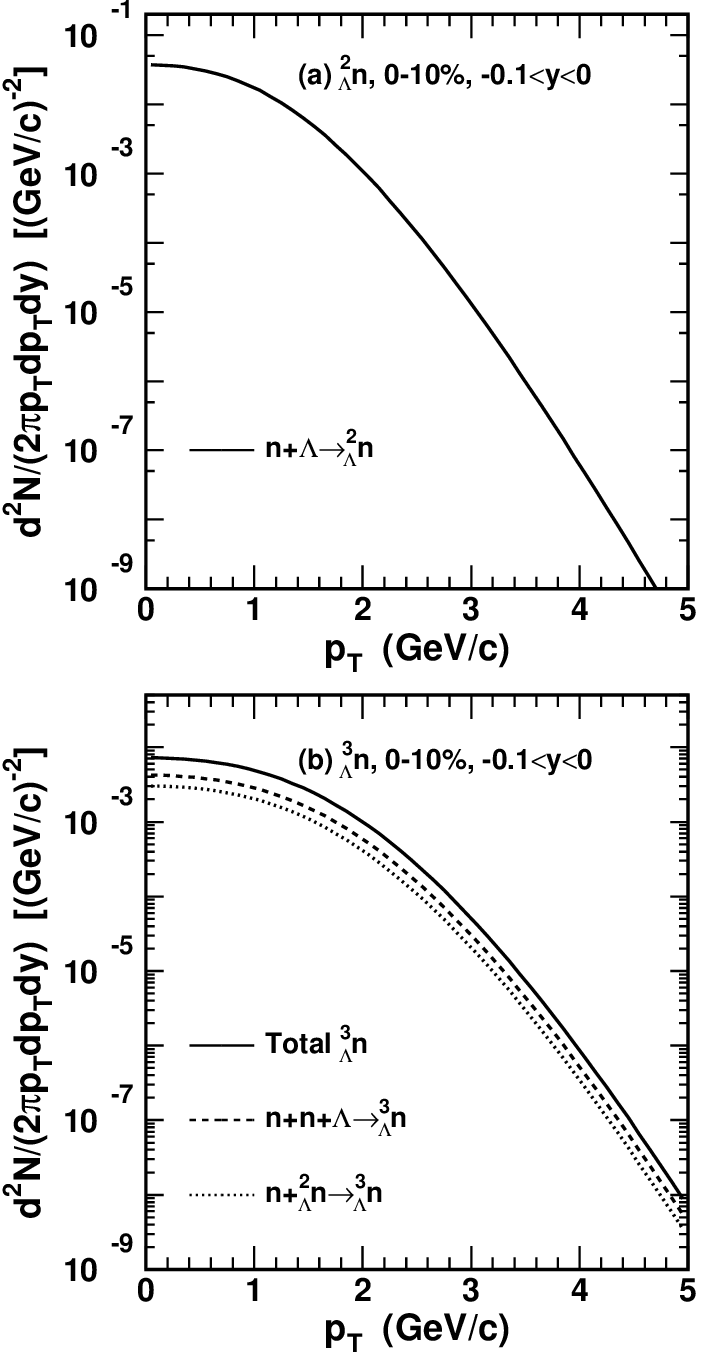}\\
 \caption{Predictions of invariant $p_T$ spectra of (a) $^2_{\Lambda}n$ and (b) $^3_{\Lambda}n$ at midrapidity in central Au-Au collisions at $\sqrt{s_{NN}}=3$ GeV.}
 \label{fig:nLnnLpT}
\end{figure}

Based on the hypothesis of their existences, we predict productions of $^2_{\Lambda}n$ and $^3_{\Lambda}n$ with Eqs.~(\ref{eq:pt-Hj2h}) and (\ref{eq:pt-Hj3h}).
We compute the invariant $p_T$ distributions of $^2_{\Lambda}n$ and $^3_{\Lambda}n$ in the rapidity interval $-0.1<y<0$ in the $0-10$\% centrality in Au-Au collisions at $\sqrt{s_{NN}}=3$ GeV.
An approximate size of $^2_{\Lambda}n$ is evaluated via $RMS_{^2_{\Lambda}n}=1/\sqrt{4\mu B_{\Lambda}}$~\cite{Bertulani:2022vad},
where the reduced mass $\mu=\frac{m_n m_{\Lambda}}{m_n+m_{\Lambda}}$ and the binding energy $B_{\Lambda}$ is 4.052 MeV~\cite{Rappold:2013fic}.
The root-mean-square radius of $^3_{\Lambda}n$ is adopted to be 2.0 fm~\cite{Hildenbrand:2019sgp}.
The spin of $^2_{\Lambda}n$ is adopted to be 0 as the spin-singlet interaction for $n\Lambda$ system is stronger than the spin-triplet interaction~\cite{Gal:2016boi}.
The spin of $^3_{\Lambda}n$ is adopted to be 1/2, the same as $^3_{\Lambda}$H.
The solid line in Fig.~\ref{fig:nLnnLpT} (a) is the result of $^2_{\Lambda}n$ produced via the process $n+\Lambda \rightarrow ^2_{\Lambda}n$.
The dashed line in Fig.~\ref{fig:nLnnLpT} (b) is the result of $^3_{\Lambda}n$ from the channel $n+n+\Lambda \rightarrow ^3_{\Lambda}n$ and the dotted line is that from $n+^2_{\Lambda}n \rightarrow ^3_{\Lambda}n$.
The solid line in Fig.~\ref{fig:nLnnLpT} (b) is the total result of the $n+n+\Lambda$ coalescence plus $n+^2_{\Lambda}n$ coalescence.

\begin{table}[!htb]
\begin{center}
\caption{Predictions of $dN/dy$ and $\langle p_T\rangle$ for $^2_{\Lambda}n$ and $^3_{\Lambda}n$ in the rapidity interval $-0.1<y<0$ in the $0-10$\% centrality in Au-Au collisions at $\sqrt{s_{NN}}=3$ GeV. } 
\label{table:dNdyavept-nLnnL}
\begin{tabular}{@{\extracolsep{\fill}}cccc@{\extracolsep{\fill}}}
\toprule
     &Coal-channel                       &$dN/dy$                               &$\langle p_T\rangle$ \\
\hline
       $^2_{\Lambda}n$
 &$n+\Lambda$                         &$1.438\times10^{-1}$             &$0.945$    \\
\hline
\multirow{3}{*}{$^3_{\Lambda}n$}
 &$n+n+\Lambda$                     &$2.808\times10^{-2}$             &$1.224$    \\
 &$n+^2_{\Lambda}n$              &$1.995\times10^{-2}$             &$1.218$    \\
 &Total                                      &$4.803\times10^{-2}$             &$1.221$    \\
\botrule
\end{tabular}
\end{center}
\end{table}

Table \ref{table:dNdyavept-nLnnL} shows predictions of yield rapidity densities $dN/dy$ and averaged transverse momenta $\langle p_T\rangle$ of $^2_{\Lambda}n$ and $^3_{\Lambda}n$ in the rapidity interval $-0.1<y<0$ in the $0-10$\% centrality in Au-Au collisions at $\sqrt{s_{NN}}=3$ GeV.
The much smaller $dN/dy$ of $^2_{\Lambda}n$ than that of the deuteron in Ref.~\cite{Wang:2022hja} comes from the seriously suppressed production of $\Lambda$ compared to the nucleon.
The enhanced $dN/dy$ of $^3_{\Lambda}n$ compared to the $^3_{\Lambda}$H closely relates with two factors.
One is the neutron surplus from the net nucleons in the colliding Au nuclei, and the other is the relatively small size of $^3_{\Lambda}n$ than $^3_{\Lambda}$H.
Our predicted multiplicities for both $^2_{\Lambda}n$ and $^3_{\Lambda}n$ locate in the regions given in Ref.~\cite{Buyukcizmeci:2024gpb}.

\subsection{Productions of $^3_{\Lambda}$H, $^4_{\Lambda}$H and $^4_{\Lambda}$He with $^2_{\Lambda}n$ and $^3_{\Lambda}n$}

Including the existences of $^2_{\Lambda}n$ and $^3_{\Lambda}n$, we re-investigate the productions of $^3_{\Lambda}$H and $^4_{\Lambda}$H and also give predictions of $^4_{\Lambda}$He.
Besides those coalescence channels shown in Fig.~\ref{fig:hyt34pT-NonLNonnL}, the additional ones contributed to the productions of $^3_{\Lambda}$H and $^4_{\Lambda}$H are as follows
{\setlength\arraycolsep{0.2pt}
\begin{eqnarray}
  && p+^2_{\Lambda}n \rightarrow ^3_{\Lambda}\text{H},  \\
  && p+n+^2_{\Lambda}n \rightarrow ^4_{\Lambda}\text{H},  \label{eq:pnnLhyt4} \\  
  && d+^2_{\Lambda}n \rightarrow ^4_{\Lambda}\text{H},  \\
  && p+^3_{\Lambda}n \rightarrow ^4_{\Lambda}\text{H}.
\end{eqnarray} }%
For $^4_{\Lambda}$He, there are four different coalescence channels in all. They are
{\setlength\arraycolsep{0.2pt}
\begin{eqnarray}
  && p+p+n+\Lambda \rightarrow ^4_{\Lambda}\text{He},  \\
  && p+d+\Lambda \rightarrow ^4_{\Lambda}\text{He},  \\
  && ^3\text{He}+\Lambda \rightarrow ^4_{\Lambda}\text{He},  \\
  && p+p+^2_{\Lambda}n \rightarrow ^4_{\Lambda}\text{He}.  \label{eq:ppnLhyt4}
\end{eqnarray} }%
Here we want to re-emphasize that $^3_{\Lambda}$H cannot take part in the formations of $^4_{\Lambda}$H and $^4_{\Lambda}$He.
This is because $^3_{\Lambda}$H is likely to be formed after $^4_{\Lambda}$H and $^4_{\Lambda}$He due to its relatively loosely-bound structure~\cite{Zhang:2018euf}, as discussed at the beginning of Sec.~\ref{IIIB}.

\begin{figure}[htbp]
\centering
 \includegraphics[width=0.95\linewidth]{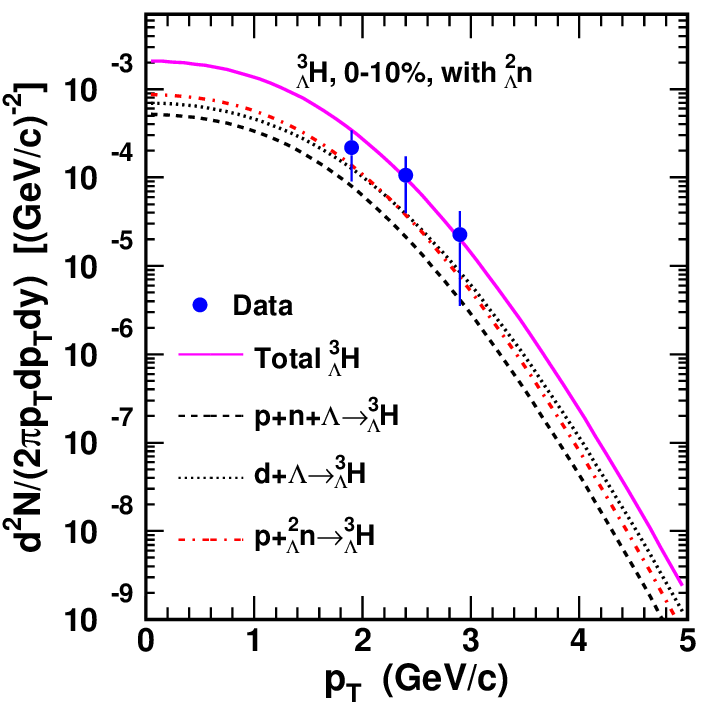}\\
 \caption{Invariant $p_T$ spectra of $^3_{\Lambda}$H including $^2_{\Lambda}n$ contribution at midrapidity in central Au-Au collisions at $\sqrt{s_{NN}}=3$ GeV. Filled circles are experimental data~\cite{STAR:2021orx}, and different lines are the theoretical results.}
 \label{fig:hytpTnew}
\end{figure}

Fig.~\ref{fig:hytpTnew} shows the invariant $p_T$ spectra of $^3_{\Lambda}$H at midrapidity in central Au-Au collisions at $\sqrt{s_{NN}}=3$ GeV. 
Filled circles with error bars are experimental data~\cite{STAR:2021orx}.
The dashed line and the dotted line denote contributions from $p+n+\Lambda \rightarrow ^3_{\Lambda}$H and $d+\Lambda \rightarrow ^3_{\Lambda}$H, respectively, the same as in Fig.~\ref{fig:hyt34pT-NonLNonnL} (a).
The dashed-dotted line is the result from $p+^2_{\Lambda}n \rightarrow ^3_{\Lambda}\text{H}$.
The solid line is the total result of the above three coalescence channels, which also reproduces the data within experimental uncertainties.

\begin{table}[!htb]
\begin{center}
\caption{Yield rapidity densities $dN/dy$ and averaged transverse momenta $\langle p_T\rangle$ of $^3_{\Lambda}$H in the rapidity interval $-0.1<y<0$ in the $0-10$\% centrality in Au-Au collisions at $\sqrt{s_{NN}}=3$ GeV. } 
\label{tab:dNdyavept-hyt3}
\begin{tabular}{@{\extracolsep{\fill}}cccc@{\extracolsep{\fill}}}
\toprule
     &Coal-channel                       &$dN/dy$                    &$\langle p_T\rangle$ \\
\hline
\multirow{4}{*}{$^3_{\Lambda}$H}
 &$p+n+\Lambda$                    &$0.323\times10^{-2}$             &$1.191$    \\
 &$d+\Lambda$                        &$0.467\times10^{-2}$             &$1.246$    \\
 &$p+^2_{\Lambda}n$             &$0.550\times10^{-2}$             &$1.200$    \\
 &Total                                     &$1.340\times10^{-2}$             &$1.214$    \\
\botrule
\end{tabular}
\end{center}
\end{table}

Table \ref{tab:dNdyavept-hyt3} shows yield rapidity densities $dN/dy$ and averaged transverse momenta $\langle p_T\rangle$ of $^3_{\Lambda}$H from different channels at midrapidity in central Au-Au collisions at $\sqrt{s_{NN}}=3$ GeV.
The total result of $dN/dy$ including $^2_{\Lambda}n$ existence is 18\% higher than the central value of the data, a little better than the result of 30\% lower in Table \ref{table:dNdyavept-hyt34} where excludes $^2_{\Lambda}n$ contribution.

\begin{figure*}[htbp]
\centering
 \includegraphics[width=0.99\linewidth]{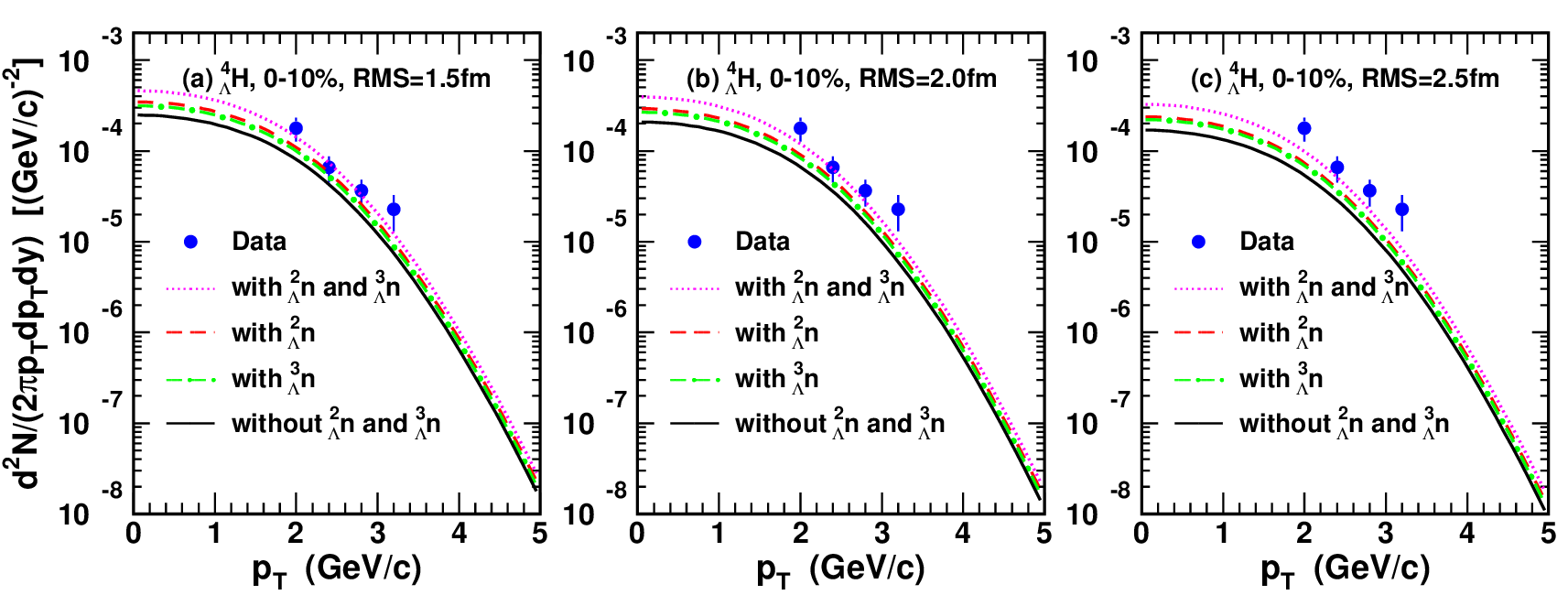}\\
 \caption{Invariant $p_T$ spectra of $^4_{\Lambda}$H at (a) RMS=1.5fm, (b) RMS=2.0fm and (c) RMS=2.5fm at midrapidity in central Au-Au collisions at $\sqrt{s_{NN}}=3$ GeV. Filled circles are experimental data~\cite{STAR:2021orx}, and different lines are the theoretical results.}
 \label{fig:hytfpTnew}
\end{figure*}

Fig.~\ref{fig:hytfpTnew} shows the invariant $p_T$ spectra of $^4_{\Lambda}$H considering $^2_{\Lambda}n$ and $^3_{\Lambda}n$ contributions.
Filled circles with error bars are experimental data~\cite{STAR:2021orx}.
To display the influences of $^2_{\Lambda}n$ and $^3_{\Lambda}n$ on the $^4_{\Lambda}$H production in detail, we give the results at four different cases.
They are including $^2_{\Lambda}n$ production only, including $^3_{\Lambda}n$ production only, including both $^2_{\Lambda}n$ and $^3_{\Lambda}n$, and without $^2_{\Lambda}n$ and $^3_{\Lambda}n$, respectively.
They are presented by dashed lines, dashed-dotted lines, dotted lines, and solid lines, respectively.
Both $^2_{\Lambda}n$ and $^3_{\Lambda}n$ enhance the $^4_{\Lambda}$H production, and the enhanced degrees are roughly same.
To study the influence of the size of $^4_{\Lambda}$H on its production, we also present results at the root-mean-square radius $RMS=1.5,~2.0,~2.5$ fm in Fig.~\ref{fig:hytfpTnew} (a), (b) and (c), respectively.
From Fig.~\ref{fig:hytfpTnew}, one can see that the $^4_{\Lambda}$H production is more suppressed at larger $RMS$ values.
Including contributions from $^2_{\Lambda}n$ and $^3_{\Lambda}n$, the theoretical results approach the experimental data much better but still underestimate the data.
Such residual underestimation may be the room left for decays of excited hypernuclei.

\begin{figure*}[htbp]
\centering
 \includegraphics[width=0.99\linewidth]{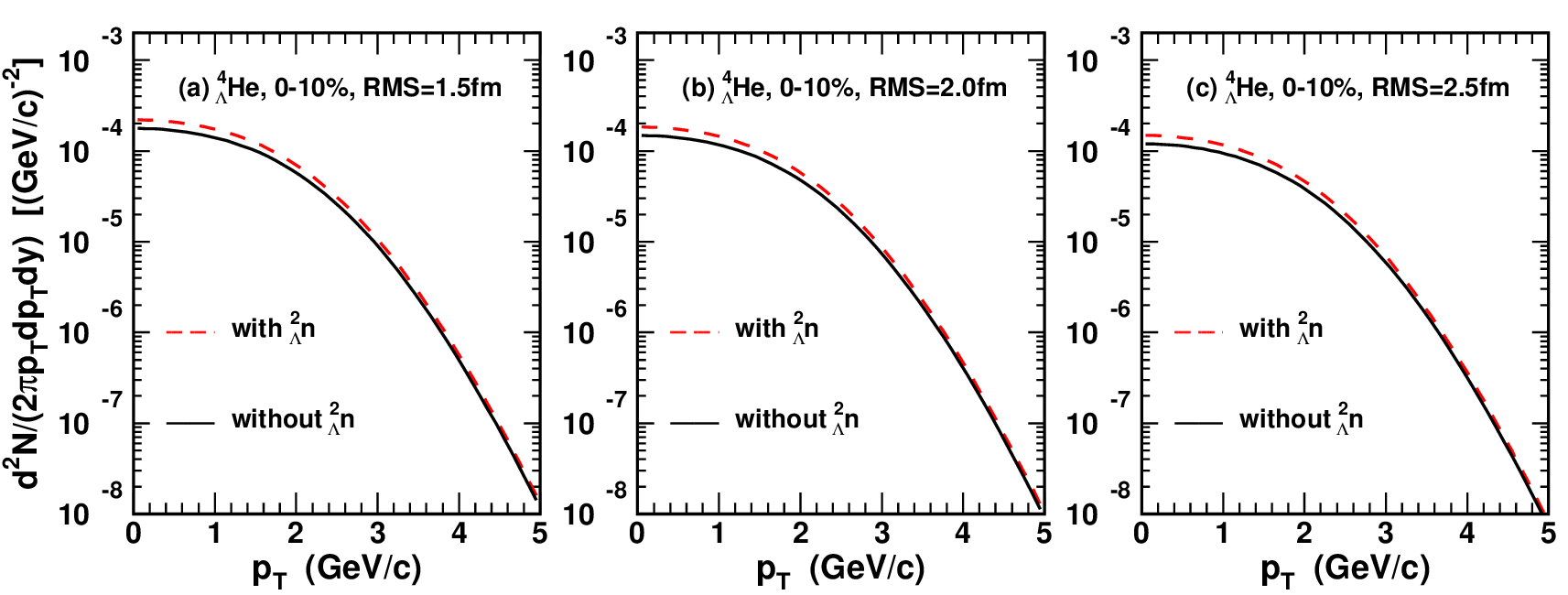}\\
 \caption{Predictions of invariant $p_T$ spectra of $^4_{\Lambda}$He at (a) RMS=1.5fm, (b) RMS=2.0fm and (c) RMS=2.5fm at midrapidity in central Au-Au collisions at $\sqrt{s_{NN}}=3$ GeV.}
 \label{fig:HeLfpTnew}
\end{figure*}

Fig.~\ref{fig:HeLfpTnew} shows the predictions of invariant $p_T$ spectra of $^4_{\Lambda}$He.
We adopt the root-mean-square radius of $^4_{\Lambda}$He to be 1.5, 2.0, 2.5 fm, the same as $^4_{\Lambda}$H~\cite{Nemura:1999qp}.
As $^3_{\Lambda}n$ does not participate in $^4_{\Lambda}$He formation, we give results at two different cases, with and without $^2_{\Lambda}n$ production.
They are presented by dashed lines and solid lines, respectively.
Future measurements of $^4_{\Lambda}$He can help to test $^2_{\Lambda}n$ contributions.

\begin{table*}[!htb]
\begin{center}
\caption{Yield rapidity densities $dN/dy$ and averaged transverse momenta $\langle p_T\rangle$ of $^4_{\Lambda}$H and $^4_{\Lambda}$He from different coalescence channels at midrapidity in central Au-Au collisions at $\sqrt{s_{NN}}=3$ GeV.} 
\label{tab:dNdyavept-hyt4HeL4}
\begin{tabular}{@{\extracolsep{\fill}}cccccccccc@{\extracolsep{\fill}}}
\toprule
     &\multirow{2}{*}{Coal-channel}     &\multicolumn{3}{@{}c@{}}{$dN/dy$}          &$~~~$          &\multicolumn{3}{@{}c@{}}{$\langle p_T\rangle$} \\
\cline{3-5}  \cline{7-9} 
                          &                               &Theory-1.5                    &Theory-2.0                  &Theory-2.5                                &      &Theory-1.5    &Theory-2.0     &Theory-2.5      \\
\hline
\multirow{8}{*}{$^4_{\Lambda}$H}
 &$p+n+n+\Lambda$                          &$0.982\times10^{-3}$   &$0.764\times10^{-3}$   &$0.565\times10^{-3}$                 &          &$1.483$        &$1.475$        &$1.467$    \\
 &$n+d+\Lambda$                              &$0.511\times10^{-3}$   &$0.423\times10^{-3}$    &$0.337\times10^{-3}$                &          &$1.520$        &$1.514$        &$1.507$    \\
 &$t+\Lambda$                                   &$1.134\times10^{-3}$   &$1.001\times10^{-3}$    &$0.864\times10^{-3}$                &          &$1.571$        &$1.566$        &$1.562$    \\
 &$p+n+^2_{\Lambda}n$                   &$0.570\times10^{-3}$   &$0.471\times10^{-3}$   &$0.374\times10^{-3}$                 &          &$1.473$        &$1.468$        &$1.462$    \\
 &$d+^2_{\Lambda}n$                       &$0.404\times10^{-3}$   &$0.350\times10^{-3}$   &$0.297\times10^{-3}$                 &          &$1.509$        &$1.505$        &$1.501$    \\
 &$p+^3_{\Lambda}n$                      &$1.114\times10^{-3}$    &$0.991\times10^{-3}$   &$0.863\times10^{-3}$                 &          &$1.471$        &$1.468$        &$1.464$    \\
 &Total                                              &$4.715\times10^{-3}$    &$4.000\times10^{-3}$  &$3.300\times10^{-3}$                  &          &$1.506$        &$1.502$        &$1.498$    \\
\hline
\multirow{6}{*}{$^4_{\Lambda}$He}
 &$p+p+n+\Lambda$                          &$0.731\times10^{-3}$   &$0.569\times10^{-3}$   &$0.420\times10^{-3}$                 &          &$1.483$        &$1.475$        &$1.467$    \\
 &$p+d+\Lambda$                              &$0.382\times10^{-3}$   &$0.316\times10^{-3}$    &$0.252\times10^{-3}$                &          &$1.520$        &$1.514$        &$1.507$    \\
 &$^3\text{He}+\Lambda$                  &$0.766\times10^{-3}$   &$0.676\times10^{-3}$    &$0.584\times10^{-3}$                &          &$1.590$        &$1.586$        &$1.581$    \\
 &$p+p+^2_{\Lambda}n$                   &$0.421\times10^{-3}$   &$0.347\times10^{-3}$   &$0.276\times10^{-3}$                 &          &$1.473$        &$1.468$        &$1.462$    \\
 &Total                                              &$2.300\times10^{-3}$    &$1.908\times10^{-3}$  &$1.532\times10^{-3}$                  &          &$1.523$        &$1.519$        &$1.516$    \\
\botrule
\end{tabular}
\end{center}
\end{table*}

To see contribution proportions of different coalescence channels for $^4_{\Lambda}$H and $^4_{\Lambda}$He more clearly, we study yield rapidity densities $dN/dy$ and averaged transverse momenta $\langle p_T\rangle$ from different channels at midrapidity in central Au-Au collisions at $\sqrt{s_{NN}}=3$ GeV.
In Table \ref{tab:dNdyavept-hyt4HeL4}, Theory-1.5, Theory-2.0 and Theory-2.5 present results of $^4_{\Lambda}$H and $^4_{\Lambda}$He at $RMS=1.5,~2.0,~2.5$ fm, respectively.
Differences of $\langle p_T\rangle$ from different coalescence channels for both $^4_{\Lambda}$H and $^4_{\Lambda}$He are less than 7\%. 
For $dN/dy$, contribution proportions from different coalescence channels are very different.
So $dN/dy$ is more powerful to justify different production channels of hypernuclei than $\langle p_T\rangle$.
 
\subsection{Production asymmetry of $^4_{\Lambda}$H and $^4_{\Lambda}$He}

As the measurements of the binding energies of $^4_{\Lambda}$H and $^4_{\Lambda}$He are consistent with each other~\cite{STAR:2022zrf} and their root-mean-square radii calculated theoretically are the same~\cite{Nemura:1999qp}, the production asymmetry of these two isobar hypernuclei mainly comes from the isospin asymmetry.
The surplus neutrons compared to protons from the colliding Au nuclei can enhance the $^4_{\Lambda}$H production more extensively than $^4_{\Lambda}$He.
The existence of $^2_{\Lambda}n$ and $^3_{\Lambda}n$ can further enlarge the difference of $^4_{\Lambda}$H and $^4_{\Lambda}$He, as shown in Eqs. (\ref{eq:pnnLhyt4}-\ref{eq:ppnLhyt4}).
Therefore, whether $^2_{\Lambda}n$ and $^3_{\Lambda}n$ exist or not will give different production asymmetries of $^4_{\Lambda}$H and $^4_{\Lambda}$He.
We propose two yield ratios $^4_{\Lambda}\text{He}/^4_{\Lambda}\text{H}$ and $(^4_{\Lambda}\text{H}-^4_{\Lambda}\text{He})/(^4_{\Lambda}\text{H}+^4_{\Lambda}\text{He})$ as deputies of the production asymmetry between $^4_{\Lambda}$H and $^4_{\Lambda}$He.
If $^4_{\Lambda}$H and $^4_{\Lambda}$He are formed symmetrically, the above two ratios are equal to 1 and 0, respectively.
Their deviations from 1 and 0 characterize the asymmetry degree.

\begin{table*}[!htb]
\begin{center}
\caption{Yield ratios of $^4_{\Lambda}$H and $^4_{\Lambda}$He at midrapidity in central Au-Au collisions at $\sqrt{s_{NN}}=3$ GeV.} 
\label{tab:ratios-hyt4HeL4}
\begin{tabular}{@{\extracolsep{\fill}}cccccccccc@{\extracolsep{\fill}}}
\toprule
 &RMS (fm)    &$~~~~$     &neither $^2_{\Lambda}n$ nor $^3_{\Lambda}n$  &$~~$      &with $^2_{\Lambda}n$ &$~~$        &with $^3_{\Lambda}n$   &$~~$        &with $^2_{\Lambda}n$ and $^3_{\Lambda}n$     \\
\hline
\multirow{3}{*}{$\frac{^4_{\Lambda}\text{He}}{^4_{\Lambda}\text{H}}$}
 &$1.5$                 &                 &$0.715$  &                                                                         &$0.639$ &                                   &$0.573$ &                                       &$0.488$    \\
 &$2.0$                 &                 &$0.713$  &                                                                         &$0.634$ &                                   &$0.564$ &                                       &$0.477$    \\
 &$2.5$                 &                 &$0.711$  &                                                                         &$0.629$ &                                   &$0.553$ &                                       &$0.464$    \\
\hline
\multirow{3}{*}{$\frac{^4_{\Lambda}\text{H}-^4_{\Lambda}\text{He}}{^4_{\Lambda}\text{H}+^4_{\Lambda}\text{He}}$}
 &$1.5$               &                 &$0.166$  &                                                                         &$0.221$ &                                   &$0.271$ &                                       &$0.344$    \\
 &$2.0$               &                 &$0.167$  &                                                                         &$0.224$ &                                   &$0.279$ &                                       &$0.354$    \\
 &$2.5$               &                 &$0.169$  &                                                                         &$0.228$ &                                   &$0.288$ &                                       &$0.366$    \\
\botrule
\end{tabular}
\end{center}
\end{table*}

We first analyze the production asymmetry of $^4_{\Lambda}$H and $^4_{\Lambda}$He formed in specific coalescence channels.
For $^4_{\Lambda}$H and $^4_{\Lambda}$He formed via four-body coalescence, with Eq.~(\ref{eq:pt-Hj4h}) we approximately have the $p_T$-integrated yield ratios
{\setlength\arraycolsep{0.2pt}
\begin{eqnarray}
 &&\frac{^4_{\Lambda}\text{He}}{^4_{\Lambda}\text{H}}  = \frac{ N_p }{N_n} = \frac{ 1 }{Z_{np}} = 0.746,    \label{eq:ratio1-4body}  \\
 &&\frac{^4_{\Lambda}\text{H}-^4_{\Lambda}\text{He}}{^4_{\Lambda}\text{H}+^4_{\Lambda}\text{He}} = \frac{ N_n-N_p }{N_n+N_p} = \frac{ Z_{np}-1 }{ Z_{np}+1 } = 0.145.    \label{eq:ratio2-4body}
\end{eqnarray} }%
Eqs.~(\ref{eq:ratio1-4body}) and (\ref{eq:ratio2-4body}) show that the production asymmetry between $^4_{\Lambda}$H and $^4_{\Lambda}$He formed via four-body coalescence closely relates with the yield density asymmetry of the neutron and the proton $Z_{np}$.
For $^4_{\Lambda}$H and $^4_{\Lambda}$He formed via three-body coalescence, their asymmetry is the same as that produced via four-body coalescence.

For $^4_{\Lambda}$H formed via $t+\Lambda$ and $^4_{\Lambda}$He formed via $^3\text{He}+\Lambda$, with Eq.~(\ref{eq:pt-Hj2h}) we approximately have the $p_T$-integrated yield ratios
{\setlength\arraycolsep{0.2pt}
\begin{eqnarray}
 &&\frac{^4_{\Lambda}\text{He}}{^4_{\Lambda}\text{H}}  = \frac{ N_{^3\text{He}} }{N_t} 
= 0.687,   \label{eq:ratio1-2body}   \\
 &&\frac{^4_{\Lambda}\text{H}-^4_{\Lambda}\text{He}}{^4_{\Lambda}\text{H}+^4_{\Lambda}\text{He}} = \frac{ N_t- N_{^3\text{He}} }{N_t+ N_{^3\text{He}}} = 0.186.  \label{eq:ratio2-2body}
\end{eqnarray} }%
The last equality signs in the above two equations are evaluated with the central values of the experimental data of $t$ and $^3\text{He}$ in Ref.~\cite{STAR:2023uxk}.
For such two-body coalescence channels, the production asymmetry between $^4_{\Lambda}$H and $^4_{\Lambda}$He is equal to the production asymmetry between $t$ and $^3$He.
The slightly larger asymmetry in Eqs.~(\ref{eq:ratio1-2body}) and (\ref{eq:ratio2-2body}), i.e., smaller $^4_{\Lambda}\text{He}/^4_{\Lambda}\text{H}$ and larger $(^4_{\Lambda}\text{H}-^4_{\Lambda}\text{He})/(^4_{\Lambda}\text{H}+^4_{\Lambda}\text{He})$ compared to those in Eqs.~(\ref{eq:ratio1-4body}) and (\ref{eq:ratio2-4body}), could come from the different sizes of $t$ and $^3$He. 
The slightly larger size of $^3$He than $t$ suppresses its production stronger in the coalescence mechanism, which is transmitted to $^4_{\Lambda}$He via two-body coalescence.

For two-body coalescence channels with $^2_{\Lambda}n$ or $^3_{\Lambda}n$ participation, only $^4_{\Lambda}$H can be formed, so
{\setlength\arraycolsep{0.2pt}
\begin{eqnarray}
 &&\frac{^4_{\Lambda}\text{He}}{^4_{\Lambda}\text{H}}  = 0,     \\
 &&\frac{^4_{\Lambda}\text{H}-^4_{\Lambda}\text{He}}{^4_{\Lambda}\text{H}+^4_{\Lambda}\text{He}} = 1. 
\end{eqnarray} }%
In these two-body channels, the production asymmetry between $^4_{\Lambda}$H and $^4_{\Lambda}$He reaches its maximum.
From the above analysis, one can see that different coalescence channels contribute different production asymmetries of $^4_{\Lambda}$H and $^4_{\Lambda}$He.

For the production asymmetry of total $^4_{\Lambda}$H and $^4_{\Lambda}$He counted all possible coalescence channels, we give the numerical results.
Table \ref{tab:ratios-hyt4HeL4} shows the calculated results for four different cases.
The first is that neither $^2_{\Lambda}n$ nor $^3_{\Lambda}n$ exists, which gives the smallest production asymmetry for the total $^4_{\Lambda}$H and $^4_{\Lambda}$He.
The second is only $^2_{\Lambda}n$ exists and the third is only $^3_{\Lambda}n$ exists.
The last is both $^2_{\Lambda}n$ and $^3_{\Lambda}n$ exist, which gives the largest production asymmetry.
We also consider the production asymmetry of $^4_{\Lambda}$H and $^4_{\Lambda}$He at different $RMS$ values, which shows that the absolute size leads to an insignificant difference in these ratios.
Whether including $^2_{\Lambda}n$ and $^3_{\Lambda}n$ productions or not gives very different results.
This further tells one that future measurements of these two ratios can help to judge the existences of $^2_{\Lambda}n$ and $^3_{\Lambda}n$.
We provide a new way to give existence constraints of the possible neutron-$\Lambda$ bound states $^2_{\Lambda}n$ and $^3_{\Lambda}n$, via productions of hypernuclei in relativistic heavy ion collisions.

\section{summary}    \label{summary}

In the coalescence mechanism, we studied the productions of various species of $\Lambda$-hypernuclei in different coalescence processes in relativistic heavy ion collisions. 
We first extended the coalescence model to the strange sector, including not only nucleon$+\Lambda$ coalescence but also nucleus+nucleon($\Lambda$) coalescence leading to productions of hypernuclei. 
We then applied the extended coalescence model to central Au-Au collisions at $\sqrt{s_{NN}}=3$ GeV to simultaneously investigate the $p_T$ spectra of $^3_{\Lambda}$H, $^4_{\Lambda}$H and $^4_{\Lambda}$He at midrapidity area.
We presented the $p_T$ distributions, the yield rapidity densities $dN/dy$ and the averaged transverse momenta $\langle p_T \rangle$ from different coalescence channels for $^3_{\Lambda}$H, $^4_{\Lambda}$H and $^4_{\Lambda}$He.
We found that yield densities from different coalescence sources for a specific kind of hypernuclei were very different, but averaged transverse momenta were almost the same.

Comparing our theoretical results to the available data published by the STAR collaboration, we found that the possible neutron-$\Lambda$ bound states $^2_{\Lambda}n$ and $^3_{\Lambda}n$ may take part in the coalescence processes to form hypernuclei with larger mass numbers.
We further predicted the production asymmetry of $^4_{\Lambda}$H and $^4_{\Lambda}$He represented by yield ratios $^4_{\Lambda}\text{He}/^4_{\Lambda}\text{H}$ and $(^4_{\Lambda}\text{H}-^4_{\Lambda}\text{He})/(^4_{\Lambda}\text{H}+^4_{\Lambda}\text{He})$, which were different in different coalescence channels.
Such asymmetric yield ratios also behaved differently in cases with or without productions of $^2_{\Lambda}n$ and $^3_{\Lambda}n$. 
Comparisons with future measurements of these ratios can help shed light on the existence constraints of $^2_{\Lambda}n$ and $^3_{\Lambda}n$.

\section*{Acknowledgements}

We thank Yi Lu for helpful discussions.
This work was supported in part by the National Natural Science Foundation of China under Grants No. 12175115 and No. 12375074, and the Doctoral Specialized Project of Nanyang Normal University under Grant No. 2025ZX006.

\bibliographystyle{apsrev4-1}
\bibliography{myref}

\end{document}